\newif\ifshortver
\shortverfalse

\ifshortver
\documentclass[conference,a4paper]{IEEEtran}
\else
\documentclass[english,onecolumn]{IEEEtran}
\fi

\IEEEoverridecommandlockouts
\ifshortver
\fi
\usepackage{soul}
\usepackage[utf8]{inputenc} 
\usepackage[T1]{fontenc}
\usepackage{url}              % provides \url{...}
\usepackage{cite}             % improves presentation of citations

\usepackage[cmex10]{amsmath}  % Use the [cmex10] option to ensure complicance
\usepackage{mleftright}       % fix to wrong spacing of \left-,
\mleftright                   % \middle- \right-commands 

\usepackage{graphicx}         % provides \includegraphics{...} to
\usepackage{caption}
\usepackage{booktabs}         % fixes poor spacing in tables and
\usepackage{amssymb}
\usepackage{esint}

\makeatletter
\def\useAmsThmOrIEEE{0}
\ifnum\useAmsThmOrIEEE=1
\usepackage{amsthm}
\theoremstyle{plain}
\newtheorem{thm}{\protect\theoremname}
\theoremstyle{definition}
\newtheorem{defn}[thm]{\protect\definitionname}
\theoremstyle{plain}
\newtheorem{prop}[thm]{\protect\propositionname}
\theoremstyle{plain}
\newtheorem{lem}[thm]{\protect\lemmaname}
\theoremstyle{plain}
\newtheorem{cor}[thm]{\protect\corollaryname}
\theoremstyle{definition}
\newtheorem{example}[thm]{\protect\examplename}
\theoremstyle{definition}
\newtheorem{rem}[thm]{\protect\remarkname}
\else
\newtheorem{thm}{\protect\theoremname}
\newtheorem{defn}[thm]{\protect\definitionname}
\newtheorem{prop}[thm]{\protect\propositionname}
\newtheorem{lem}[thm]{\protect\lemmaname}
\newtheorem{cor}[thm]{\protect\corollaryname}
\newtheorem{example}[thm]{\protect\examplename}

\fi

\usepackage{mathrsfs}
\usepackage{algorithm,algpseudocode}

\usepackage{colortbl}
\definecolor{lightgray}{rgb}{0.9,0.9,0.9}
\definecolor{lightred}{rgb}{1,0.8,0.8}
\definecolor{lightgreen}{rgb}{0.6,1,0.6}
\definecolor{lightyellow}{rgb}{1,1,0.5}
\definecolor{lightgrey}{rgb}{0.8,0.8,0.8}

\allowdisplaybreaks[1]

\makeatother

\usepackage{babel}
\providecommand{\corollaryname}{Corollary}
\providecommand{\definitionname}{Definition}
\providecommand{\lemmaname}{Lemma}
\providecommand{\propositionname}{Proposition}
\providecommand{\theoremname}{Theorem}
\providecommand{\examplename}{Example}
\providecommand{\remarkname}{Remark}

\begin{document}
\title{Vector Quantization with Error Uniformly Distributed over an Arbitrary Set}
% \author{%
%   \IEEEauthorblockN{Chih Wei Ling and Cheuk Ting Li}
% \ifshortver
% \else
% \\
% \fi
%   \IEEEauthorblockA{%
%     Department of Information Engineering, The Chinese University of Hong
% Kong\\
% Email: chihweiLing@link.cuhk.edu.hk,  ctli@ie.cuhk.edu.hk}
% \ifshortver
% \thanks{
% Some proofs are omitted due to space constraint. They can be found in~\cite{vector_arxiv}.
% }
% \fi
% }
\author{%
Chih Wei Ling and Cheuk Ting Li, \textit{Member, IEEE}
\thanks{
The work described in this paper was partially supported by an
ECS grant from the Research Grants Council of the Hong Kong Special Administrative
Region, China [Project No.: CUHK 24205621].

This paper was presented in part at the 2023 IEEE International Symposium on Information Theory (ISIT).

Chih Wei Ling and Cheuk Ting Li are with the Department of Information Engineering, The Chinese University of Hong
Kong, Hong Kong SAR of China. Email: chihweiLing@link.cuhk.edu.hk, ctli@ie.cuhk.edu.hk.
}
}
\maketitle
\begin{abstract}
For uniform scalar quantization, the error distribution is approximately a uniform distribution over an interval (which is also a 1-dimensional ball). Nevertheless, for lattice vector quantization, the error distribution is uniform not over a ball, but over the basic cell of the quantization lattice. In this paper, we construct vector quantizers with periodic properties, where the error is uniformly distributed over the $n$-ball, or any other prescribed set. We then prove upper and lower bounds on the entropy of the quantized signals.
We also discuss how our construction can be applied to give a randomized quantization scheme with a nonuniform error distribution.
\end{abstract}

\medskip{}

\section{Introduction} \label{sec:intro}

The quantization problems, and their relations to the packing and covering properties of $n$-dimensional lattices or non-lattice arrangements of points have been studied extensively in mathematical and information-theoretic literature \cite{conway2013sphere,zamir2014,gersho1979,conway1982,Toth1959SurLR,barnes1983}. One of the central questions in the study of quantization is the characterization of the distribution of the quantization error (or error distribution).
A common assumption is that the quantization error of a  uniform scalar quantizer is approximately uniformly distributed over an interval, and the quantization error of a lattice quantizer is approximately uniformly distributed over the basic cell of the quantizer~\cite{gray1998quantization}.
Bennett~\cite{bennett1948spectra} studied the spectra of quantized signals.
Furthermore, Gariby and Erez~\cite{gariby2008general} showed that there exists a sequence of lattices with proper partitions such that their quantization errors approach a desired distribution in the KL divergence sense.

Dithering is a technique for constructing randomized quantizers which ensures that the error is uniformly distributed over an interval (or, more generally, a basic cell of a lattice), independent of the source \cite{ roberts1962, schuchman1964dither, Limb1969, Jaynant1972, sripad1977necessary}.
Zamir and Feder~\cite{Zamir1996Noise} proved that the spectra of the quantization error of an optimal lattice quantizer (with or without a dither) is white, there exists a sequence of lattices which are asymptotically optimal in the normalized second moment sense (i.e., achieving Zador's bound \cite{Zador1982asymptotic,Gray2002onZador}), and the quantization errors approach a white Gaussian process in the KL divergence sense as the dimension increases.
Additionally, Kirac and Vaidyanathan \cite{kirac1996results} analyzed dithered lattice vector quantizers, and gave conditions under which the error is uniformly distributed over a basic cell of the lattice.

% Note that the resulting quantization error in the aforementioned deterministic or randomized quantization schemes, %in finite dimensions, 
% i.e., without or with dithering, follows approximately uniform or uniform distributions over certain sets (e.g., some basic cells of a lattice) under suitable assumptions.
Note that the resulting quantization error in the aforementioned quantization schemes approximately follows a uniform distribution over a basic cell of a lattice.\footnote{For deterministic quantization schemes without dithering, the error is approximately uniform over a basic cell under suitable assumptions. For dithered quantization, the error is exactly uniform over a basic cell.}
Instead of having the error distributed uniformly over a basic cell, it may be desirable to generalize it so that the error follows a uniform distribution over an arbitrary set.
Note that any nonuniform distribution can be expressed as a mixture of uniform distributions, by using the layered construction described in \cite{wilson2000layered,hegazy2022randomized}. Therefore, if we can construct a quantization scheme with an error uniform over over an arbitrary set, then we can construct a quantization scheme with an arbitrary error distribution.
% Given any prescribed nonuniform error distribution, 
% by using the slicing techniques as described in \cite{hegazy2022randomized,wilson2000layered}, which treat each level set of the distribution as a single error following a uniform distribution over the level set or any similar slicing technique, it is possible to regulate the error to follow the prescribed  distribution. 
This has applications in differential privacy and machine learning, described as follows.

In differential privacy \cite{Dwork06DP,Dwork14Book,duchi2013local}, 
a random noise following a prescribed distribution, such as a Laplace or Gaussian distribution, is added to the utility function of a dataset to ensure privacy.
Such additive noise differential privacy mechanisms are widely used in federated learning \cite{el2022differential,li2020federated} 
% and have been deployed in commercial database management systems \cite{kessler2019sap} 
to provide privacy guarantees.
% Furthermore, Duchi et al. \cite{duchi2013local}  formulated the local differential privacy problem as a calculus of variation problem, where the solutions are general channels that satisfy both the privacy and utility constraints, while Xiong et al. \cite{xiong2016randomized} provided a version of discrete memoryless channel. 
% This interpretation can be translated into the construction of a randomized quantization scheme for differential privacy, and various construction methods have been proposed \cite{shahmiri2023,hegazy2023compression,lang2022joint}.
% In this regard, Lang et al.~\cite{lang2022joint} and Shahmiri et al.~\cite{shahmiri2023} utilized a randomized quantizer with a Laplace error distribution to ensure privacy in federated learning.
Dithered quantization techniques have been applied to construct additive noise mechanisms with a small communication cost in \cite{lang2022joint,shahmiri2023,hasircioglu2023communication,hegazy2023compression,yan2023layered}.
In addition to differential privacy, controlling the distribution of the quantization error is also crucial in various machine learning tasks \cite{agustsson2020universally,flamich2020compressing,shlezinger2020uveqfed,yang2022introduction}.

The above applications motivate the study on constructing a quantization scheme with an arbitrary prescribed error distribution. 
Hegazy and Li \cite{hegazy2022randomized} designed an optimal randomized scalar quantization scheme with the length of quantization step follows a random distribution so that the quantization error is independent of the source and follows any given unimodal distribution. 
Recently, randomized quantization schemes for simulating Gaussian and Laplace noises have been explored as a means to ensure privacy in \cite{lang2022joint,shahmiri2023,hasircioglu2023communication,hegazy2023compression,yan2023layered}.

In this paper, which is the extended version of~\cite{vectquant_isit},\footnote{Compared to the short version~\cite{vectquant_isit}, this full version includes  a  quantizer with only finitely many quantization cells per basic cell in Theorem~\ref{thm:quantization_tv}, a new, sharper lower bound on the normalized entropy based on sphere packing in Theorem~\ref{thm:ub_on_ball}, a new example in Figure~\ref{fig::quantizerComparison}, the layered shift-periodic quantizer ensemble with dithering for nonuniform error distributions in Section~\ref{sec:nonunif}, and the proofs of all the results (\cite{vectquant_isit} only includes the first part of the proof of Lemma~\ref{lem:dissect} and the proof of Theorem~\ref{thm:quantization}).} we present new constructions of vector quantizers where the error can be approximately uniformly distributed over an arbitrary given set, or exactly uniformly distributed over that set when used with dithering. In contrast, conventional lattice quantizers always give an error distribution that is uniform over a basic cell~\cite{gray1998quantization}. Although the proposed quantizers are constructed using a lattice, and have the same periodic property as conventional lattice quantizers, the main difference is that the quantization cells of the proposed quantizers can take many (even infinitely many) different shapes. This allows the error to be uniformly distributed over sets with curved boundaries, such as a ball. We then prove achievability and converse bounds on the entropy of the quantized signal when the input signal is uniform over a large set (see Proposition~\ref{prop:ent_large} and Theorem~\ref{thm:quantization}).

We are particularly interested in designing quantizers with error uniformly distributed over a ball. This allows the quantization to be ``circularly symmetric'', i.e., the quantization error looks the same regardless of the direction. The simpler characterization of a ball (compared to the polyhedral basic cells) allows the error distribution to be tracked more easily. This may be useful, for example, in the gradient computation in neural networks \cite{agustsson2020universally}, since representing the location of a ball (or an ellipsoid after linear transformation) is simpler than representing a polytope. For the 2D case, we construct a quantizer with error uniform over a disk based on the hexagonal lattice, and conjecture that the hexagonal lattice is optimal. For the 3D case, we conjecture that the face-centered cubic lattice is optimal.

We remark that, since any continuous nonuniform distribution can be expressed as a mixture of uniform distributions (see \cite{hegazy2022randomized}), the quantizers proposed in this paper can be applied in a randomized manner to allow the error distribution to be any continuous (uniform/nonuniform) distribution, via a technique similar to \cite{hegazy2022randomized}. More specifically, if we allow common randomness between the encoder and the decoder, they can use the common randomness to agree on a random quantizers, and have an error distribution that is the mixture of the error distributions of the individual quantizers.
As mentioned above, this has applications in differential privacy, enabling the constructions of additive noise mechanisms. For example, the multivariate Gaussian distribution, multivariate Laplace distribution~\cite{andres2013geo},\footnote{There are multiple generalizations of the Laplace distribution to higher dimensions. Here we refer to the one used in geo-indistinguishability~\cite{andres2013geo}, with density $f(\mathbf{z}) \propto e^{- \epsilon \Vert \mathbf{z} \Vert}$.} and other elliptical distributions \cite{reimherr2019elliptical}, can be expressed as mixtures of uniform distributions over balls (or ellipsoids), making the study on quantization with uniform error distributions over balls particularly interesting.

\ifshortver
Some proofs are omitted due to space constraint. They can be found in the full version~\cite{vector_arxiv}.
\fi

\subsection*{Other Related Works}

% Dithered quantization was conceived for enhancing the perceptual effect of picture and speech coding in \cite{ roberts1962, schuchman1964dither, Limb1969, Jaynant1972}.
% A primary motivation of using dithered quantization is its possibility to produce a 
% quantization error that is independent of the source
%, if the Schuchman's condition is satisfied 
% \cite{schuchman1964dither}.
Traditionally, dithered quantization is studied under the setting where the quantizer is uniform with a fixed step size $h$, and the dither signal is uniformly distributed over the interval $(-h/2, h/2)$.
For a comprehensive survey,
%of subtractive dither as well as non-subtractive dither, 
see \cite{gray1993dithered,gray1998quantization,Lipshitz1992}.
Ziv \cite{Ziv1985} used the idea of entropy-coded quantization (a dithered quantizer followed by a entropy coding) to construct a scheme that has nearly optimal performance regardless of the input distribution.
Gray~\cite{Gray1990spectra} studied the spectra of quantized signals of dithered quantizers.
Comprehensive studies on the properties of dithered quantizers and entropy-coded quantization was performed by Zamir and Feder \cite{Zamir1992,Zamir1995RD,Zamir1996}. 
By using Wiener filtering and entropy-coded quantization with a feedback loop for noise shaping, Mashiach and Zamir designed a scheme for the nonuniform sampling of a continuous-time Gaussian source that achieves the rate-distortion function of the source \cite{Mashiach2013noiseShaping}.
Similar constructions based on entropy-coded quantization for the sampling of continuous signals were observed in \cite{Zamir1995RD,Zamir1996}.
Moreover, nonuniform dithered quantization was investigated by Akyol and Rose in \cite{Akyol2012}.

Channel simulation \cite{bennett2002entanglement} is a closely related setting in which the encoder compresses an input source $X$ into a message $M$, followed by the decoder recovering $Y$ using $M$, and the goal is to make $Y$ follow a prescribed conditional distribution given $X$.
Channel simulation was originally studied in the asymptotic setting \cite{bennett2002entanglement, winter2002compression, cuff2013distributed} where $X$ is a sequence, and was extended to the one-shot setting \cite{Harsha2007,braverman2014public,li2018universal,sfrl_trans} where $X$ is a scalar.
%Note that randomized quantization can be regarded as a special case of the channel simulation by treating the distribution of the quantization error as the prescribed conditional distribution.
One-shot channel simulation with common randomness \cite{Harsha2007,braverman2014public,sfrl_trans}, when applied to simulate an additive noise channel $Y=X+Z$, can be regarded as
a randomized quantization scheme where the quantization error distribution (i.e., the distribution of the channel noise $Z$)
can be an arbitrary prescribed distribution. 
% Nevertheless, the construction
% in \cite{sfrl_trans,li2021unified} is randomized and aperiodic, and
% does not admit a practical implementation. Also, \cite{sfrl_trans,li2021unified}
% requires the encoder and decoder to share common randomness, which
% is not necessary in this paper. 
One-shot channel simulation without
common randomness has also been studied in \cite{kumar2014exact,li2017distributed,li2018universal},
though local randomness at the decoder is still needed, making these schemes 
different from conventional quantization.

\subsection*{Notations}

Logarithms are to the base $2$, and entropy is in bits. For measurable
$A\subseteq\mathbb{R}^{n}$, write $\mu(A)$ for its Lebesgue measure.
The binary entropy function (i.e., the entropy of $\mathrm{Bern}(p)$)
is written as $H_{b}(p):=-p\log p-(1-p)\log(1-p)$. For $A,B\subseteq\mathbb{R}^{n}$, $A+B:=\{a+b:\,a\in A,b\in B\}$,
$A-B:=\{a-b:\,a\in A,b\in B\}$, and $\mathbf{G}A:=\{\mathbf{G}a:\,a\in A\}$ for $\mathbf{G} \in \mathbb{R}^{n \times n}$. We usually write $B_n := \{\mathbf{x} \in \mathbb{R}^n:\, \Vert \mathbf{x}\Vert \le 1\}$ for the closed unit $n$-ball. 
% For a function $f:\mathbb{R}^n \rightarrow \mathbb{R}$, its superlevel set is defined as
% $L_{\tau}^{+}(f):=\{\mathbf{x} \in \mathbb{R}^{n}: f(\mathbf{x}) \geq \tau\}$,
% where $\tau \in \mathbb{R}$. 
For two distributions $P,Q$ over the same set, their total variation distance is $\delta_{\mathrm{TV}}(P,Q):=\sup_{A}|P(A)-Q(A)|$, where the supremum is over measurable sets $A$. For a random variable $X$, we write $\delta_{\mathrm{TV}}(X,Q)=\delta_{\mathrm{TV}}(P,Q)$ where $P$ is the distribution of $X$.

\smallskip

\section{Shift-Periodic Quantizer}

Before we introduce our construction, we review some basic concepts in lattice quantization.
The \emph{lattice} generated by the full-rank matrix $\mathbf{G}\in\mathbb{R}^{n\times n}$ is the set $\Lambda(\mathbf{G}) :=\{\mathbf{G}\mathbf{v}:\, \mathbf{v}\in\mathbb{Z}^{n}\}$.
A \emph{basic cell} (or fundamental cell) $S\subseteq\mathbb{R}^{n}$ of the lattice generated by $\mathbf{G}\in\mathbb{R}^{n\times n}$ \cite{conway2013sphere,zamir2014}
is a bounded set where it is possible to partition the space using lattice-translated
copies of $S$, i.e.,
\[
\bigcup_{\mathbf{v}\in\mathbb{Z}^{n}}(S+\mathbf{G}\mathbf{v}) = \mathbb{R}^{n},
\]
and $S\cap(S+\mathbf{G}\mathbf{v}) = \emptyset$ for any $\mathbf{v}\in\mathbb{Z}^{n}\backslash\{\mathbf{0}\}$. 
One example is the parallelepipeds $\left\{ \mathbf{G}\mathbf{u}:\,\mathbf{u}\in[0,1)^{n}\right\}$.
For $\mathbf{x} \in \mathbb{R}^n$, we write $\mathbf{x}\; \mathrm{mod}\; S := \mathbf{x} - \mathbf{G}\mathbf{v}$, where $\mathbf{v}\in\mathbb{Z}^{n}$ satisfies $\mathbf{x} \in S+\mathbf{G}\mathbf{v}$.
% \[
% \mathbb{R}^{n}\backslash\bigcup_{\mathbf{v}\in\mathbb{Z}^{n}}(S+\mathbf{G}\mathbf{v})
% \]
% is a null set, and $S\cap(S+\mathbf{G}\mathbf{v})$ is a null
% set for any $\mathbf{v}\in\mathbb{Z}^{n}\backslash\{\mathbf{0}\}$.
% One particular type of parallelohedron is \emph{parallelepipeds} \cite{alex2005convex, grunbaum2003convex},
% i.e., sets which can be expressed in the form
% \[
% \left\{ \mathbf{G}\mathbf{u}:\,\mathbf{u}\in[0,1)^{n}\right\} \subseteq\mathbb{R}^{n}
% \]
% for some full-rank matrix $\mathbf{G}\in\mathbb{R}^{n\times n}$.

A \emph{Nyquist-$\mathbf{G}$ distribution} 
\cite{kirac1996results} 
(or a distribution with \emph{constant periodic replication} \cite{zamir2014})
is a distribution with probability density
function $f$ satisfying
\begin{equation} \label{eq:NyquistG}
\sum_{\mathbf{v}\in\mathbb{Z}^{n}}f(\mathbf{t}+\mathbf{G}\mathbf{v})=\frac{1}{|\det\mathbf{G}|}
\end{equation}
for all $\mathbf{t}\in\mathbb{R}^{n}$. 
The left-hand side of (\ref{eq:NyquistG})
% \footnote{\textcolor{blue}{Condition~(\ref{eq:NyquistG}) is also called \emph{constant periodic replication} in \cite{zamir2014}}} 
is called the \emph{periodic replication}\footnote{Notice that the periodic replication of $f$ is periodic with respect to $\Lambda(\mathbf{G})$, i.e., if $f_{\textsc{rep} \; \Lambda(G)}(\mathbf{t}):=\sum_{\mathbf{v}\in\mathbb{Z}^{n}}f(\mathbf{t}+\mathbf{G}\mathbf{v})$, then $f_{\textsc{rep} \; \Lambda(G)}(\mathbf{t}+\mathbf{G}\mathbf{v}) = f_{\textsc{rep} \; \Lambda(G)}(\mathbf{t})$, for all $\mathbf{v} \in \mathbb{Z}^n$.} of $f$ by $\Lambda(\mathbf{G})$, i.e., the sum of the function over the coset of $\mathbf{t}$ with respect to $\Lambda(\mathbf{G})$.
Equivalently, a Nyquist-$\mathbf{G}$ distribution is a distribution $f$ where the corresponding \emph{folded distribution} \cite[Equation (4.16)]{zamir2014}, i.e., 
the distribution of $\mathbf{x} \;\mathrm{mod}\; S$ (where $\mathbf{x} \sim f$ and $S$ is a basic cell), is the uniform distribution over the basic cell $S$.
One example of Nyquist-$\mathbf{G}$ distributions is the uniform distribution over a basic cell.
Refer to Appendix~\ref{sec:nyquist} for more examples.

\begin{figure}
    \centering
    \includegraphics[width=400pt]{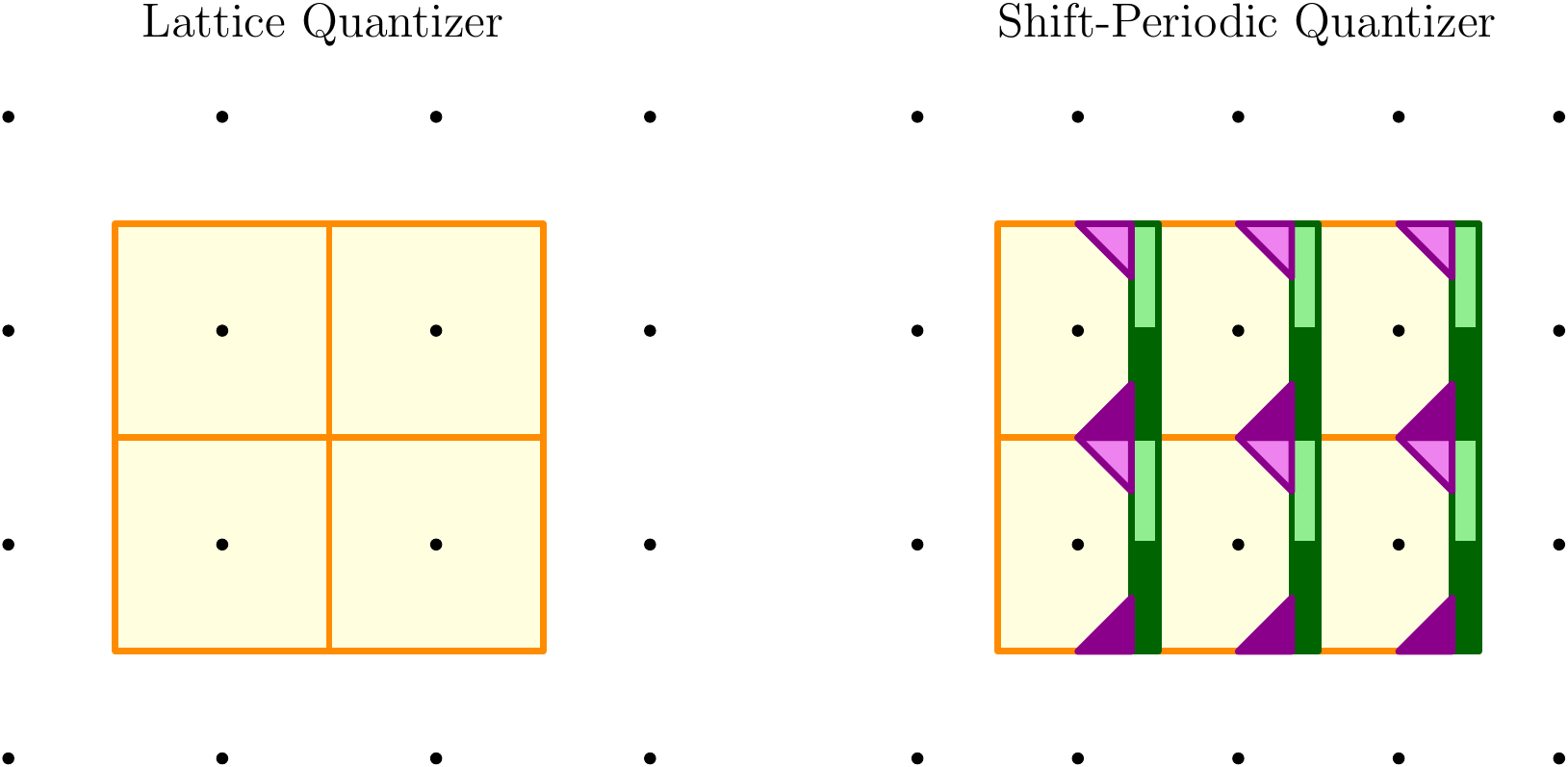}

    \vspace{20pt}
    \includegraphics[width=330pt]{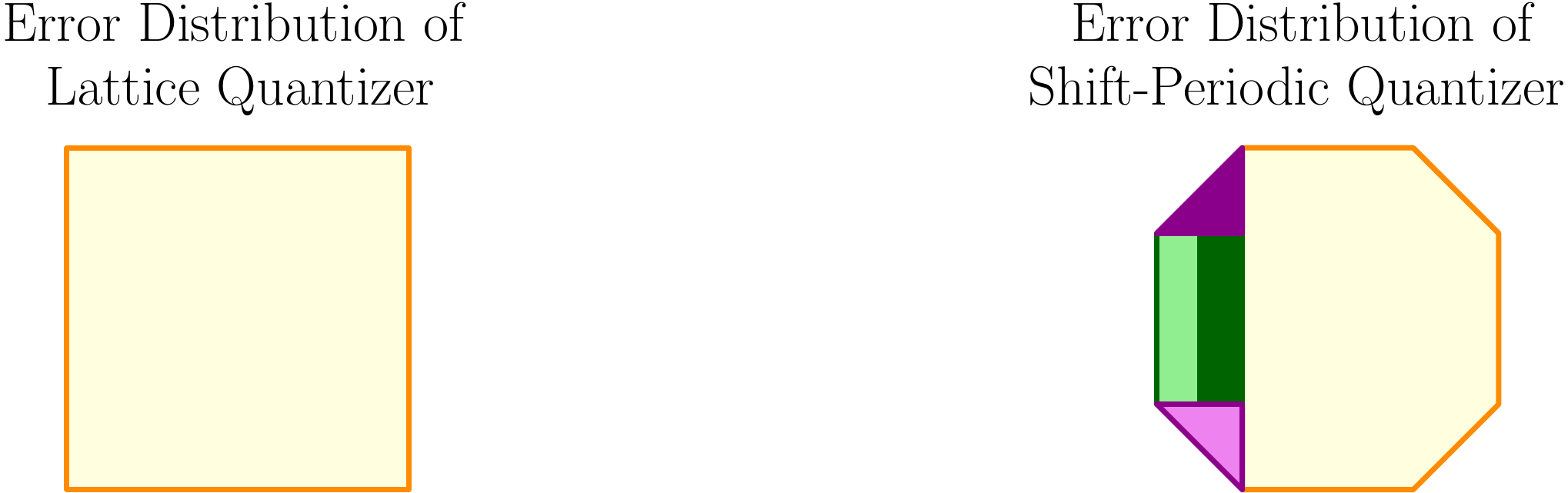}
    
    \vspace{20pt}
    \includegraphics[width=420pt]{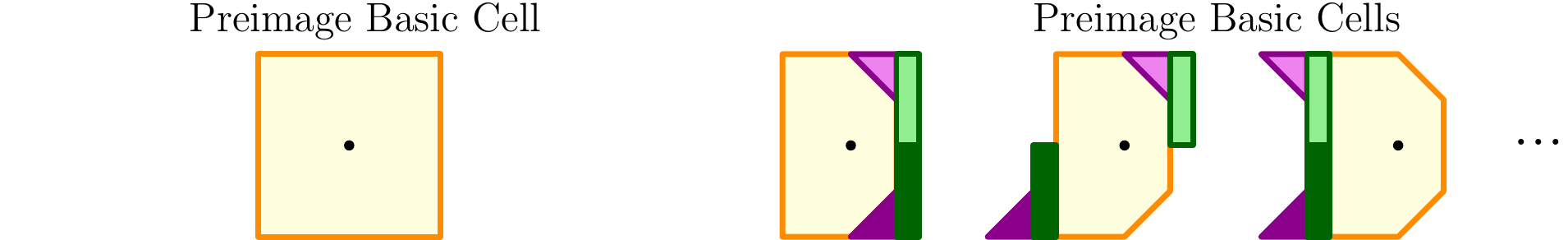}
    
    \vspace{10pt}    \caption{\label{fig:shiftperiodic}The comparison of a shift-periodic quantizer with a lattice quantizer. The top left figure shows a lattice quantizer in which the quantization cell is a square-shaped basic cell. The distribution of the error $\mathbf{x} - Q(\mathbf{x})$ (when the source $\mathbf{x}$ is  uniformly distributed over a basic cell $S$ of $\Lambda(\mathbf{G})$ or, more generally, follows a Nyquist-$\mathbf{G}$ distribution) is uniform over the same square (middle left figure).
    The bottom left figure shows that the only preimage basic cell is the square-shaped basic cell (and its shifted versions).
    %Conditional on the location of the input vector $\mathbf{x}$ (suppose $\mathbf{x}$ is in the first square, and all the content in the square is quantized to the lattice point corresponding to the first square), evidently, the error distribution is uniformly distributed over the same square as well.
    The top right figure shows a shift-periodic quantizer, where the rectangular basic cell of the lattice is divided into $5$ quantization cells of different shapes and sizes. By choosing the reconstruction points of each quantization cell, we can have an error distribution that is uniform over an octagon formed by translating the $5$ quantization cells (middle right figure).
    % Conditional on the input vector $\mathbf{x}$ (say $\mathbf{x}$ is in the first rectangle), we can find a quantizer (by first part Theorem~\ref{thm:quantization}) so that the error distribution is uniformly distributed over an octagon, which is periodic with respect to the lattice points in space.
    % \textcolor{red}{The bottom right figure shows that the (blue-shaded) rectangular basic cell and preimage basic cell (irregular shape) are different. 
    % The colored lines represent the reconstruction points as position vectors. 
    % For illustrative purposes, we have chosen to use a rectangular basic cell with the lattices as the vertices instead of the Voronoi region.}
    The bottom right figure shows some examples of preimage basic cells. Recall that a preimage basic cell is formed by the union of quantization cells, where exactly one quantization cell is chosen from each kind of quantization cells (there are $5$ kinds of quantization cells distinguished by colors).
    }
    \label{fig::quantizerComparison}
\end{figure}

In this paper, a quantizer is a measurable function $Q:\mathbb{R}^{n}\to\mathbb{R}^{n}$
where the range $Q(\mathbb{R}^{n})=\{Q(\mathbf{x}):\,\mathbf{x}\in\mathbb{R}^{n}\}$
is a countable set. Unlike conventional lattice quantizers where the quantization cells are shifted copies of the same basic cell, we only require that a quantizer is periodic with respect to the lattice with generator matrix $\mathbf{G}$. Therefore, we also consider quantizers with several (or even infinitely many) different shapes of quantization cells (e.g. Figure~\ref{fig::quantizerComparison} and  Figure~\ref{fig::constructionF}). 
This will be stated precisely in the following definition.

\smallskip
\begin{defn}
We call the quantizer $Q:\mathbb{R}^{n}\to\mathbb{R}^{n}$
\emph{shift-periodic} with generator matrix $\mathbf{G}\in\mathbb{R}^{n\times n}$
(a full-rank matrix) if 
\[
Q(\mathbf{x}+\mathbf{G}\mathbf{v})=Q(\mathbf{x})+\mathbf{G}\mathbf{v}
\]
for all $\mathbf{x}\in \mathbb{R}^n$, $\mathbf{v}\in\mathbb{Z}^{n}$ (this is equivalent to saying that the quantization error $\mathbf{x}-Q(\mathbf{x})$ is periodic over the lattice generated by $\mathbf{G}$), and the magnitude of the quantization error $\Vert \mathbf{x}-Q(\mathbf{x})\Vert$ has a finite upper bound for $\mathbf{x}\in \mathbb{R}^n$. 
% \textcolor{blue}{Note that this definition is equivalent to saying that if $A$ is a quantization cell of $Q$, then $A+\mathbf{G} \mathbf{v}$ for $\mathbf{v} \in \mathbb{Z}^{n}\backslash \{\mathbf{0}\}$ is also a quantization cell of $Q$}
% The \emph{normalized
% entropy} of a shift-periodic quantizer $Q$ is 
% \[
% \bar{H}(Q):=H\big(\mathrm{frac}\big(\mathbf{G}^{-1}Q(\mathbf{G}\mathbf{u})\big)\big)-\log|\det\mathbf{G}|,
% \]
% where $\mathbf{u}\sim\mathrm{Unif}([0,1)^{n})$, and $\mathrm{frac}(\mathbf{x})=(x_{1}-\lfloor x_{1}\rfloor,\ldots,x_{n}-\lfloor x_{n}\rfloor)$
% is the entrywise fractional part function.
\end{defn}
\smallskip
% We describe two scenarios where we can apply a shift-periodic quantization
% function.
% \begin{enumerate}
% \item ??????
% \item Dithered vector quantization
% \end{enumerate}
% ??????????????

% In Figure~\ref{fig::quantizerComparison}, we compare a shift-periodic quantizer with a lattice quantizer. 
% In Figure~\ref{fig::errorComparison}, we show the difference between the error distributions of the two quantizers.
In Figure~\ref{fig::quantizerComparison}, we present a simple example of a shift-periodic quantizer, and compare it to a lattice quantizer. While a lattice quantizer must have an error distribution uniform over a basic cell of the lattice (e.g. a square or a regular hexagon), a shift-periodic quantizer can have an error distribution uniform over other sets (e.g. an octagon in the example). The shift-periodic quantizer in the figure might be desirable over lattice quantizers if the goal is to have an error distribution approximate uniform over a circle, since an octagon is closer to a circle than a square or a hexagon. We will see later in this paper that the error distribution of a shift-periodic quantizer can be uniform over any shape. In particular, it can be exactly uniform over a circle.

% \smallskip

% \begin{figure}
%     \centering
%     \includegraphics[width=300pt]{ErrorComparison.pdf}
%     \caption{The distributions of errors $\mathbf{x} - Q(\mathbf{x})$ of a lattice quantizer and a shift-periodic quantizer. The left figure shows the distribution of the error $\mathbf{x} - Q(\mathbf{x})$ of a lattice quantizer - a uniform distribution over a square. The right figure shows the distribution of the error $\mathbf{x} - Q(\mathbf{x})$ of a shift-periodic quantizer - a uniform distribution over an octagon.}
%     \label{fig::errorComparison}
% \end{figure}
% \smallskip

In the following subsections, we discuss some basic properties of shift-periodic quantizers, and characterize the error distribution and entropy of these quantizers.

\subsection{Error Distribution}
Although the distribution of the quantization error $\mathbf{x}-Q(\mathbf{x})$ depends on the distribution of $\mathbf{x}$, there is an ``average-case'' error distribution which is given below.
\smallskip
\begin{defn}\label{def:err_dist}
The \emph{error distribution}
$\bar{f}_{Q}:\mathbb{R}^{n}\to\mathbb{R}$ of a shift-periodic quantization
function $Q$ is the probability density function of 
$\mathbf{x}-Q(\mathbf{x})$,
where $\mathbf{x}\sim\mathrm{Unif}(S)$ is uniform over a basic cell $S$. 
% \[
% \mathbf{G}\mathbf{u}-Q(\mathbf{G}\mathbf{u}),
% \]
% where $\mathbf{u}\sim\mathrm{Unif}([0,1)^{n})$. 
\end{defn}
\smallskip

Note that $\bar{f}_{Q}$ may be uniform or nonuniform. For example, for $Q:\mathbb{R}\to\mathbb{R}$, $Q(x) = \mathrm{argmin}_{v \in \mathbb{Z}} |x-v|$, $\bar{f}_Q$ is the uniform distribution $\mathrm{Unif}(-1/2,1/2)$. For $Q(x) = \mathrm{argmin}_{v \in \mathbb{Z}} (2+(-1)^v)|x-v|$ (which is shift-periodic with period $2$), $\bar{f}_Q$ is the mixture distribution $(1/4)\mathrm{Unif}(-1/4,1/4) + (3/4)\mathrm{Unif}(-3/4,3/4)$.
To show that Definition~\ref{def:err_dist} is valid, we need to prove that the error distribution is the same for all basic cells $S$. In fact, the distribution is the same whenever $\mathbf{x}$ follows a Nyquist-$\mathbf{G}$
distribution. The proof is in
\ifshortver
\cite{vector_arxiv}.
\else
Appendix~\ref{subsec:quantize_nyquist_pf}.
\fi
\smallskip
\begin{prop}\label{prop:quantize_nyquist}
Consider a shift-periodic quantizer $Q$.
For any Nyquist-$\mathbf{G}$
distribution, assuming the input vector $\mathbf{x}$ follows that
distribution, the error $\mathbf{x}-Q(\mathbf{x})$
follows the error distribution $\bar{f}_{Q}$.
\end{prop}
\smallskip
This implies that if $\mathbf{x}$ is uniformly distributed over a large set, the error distribution is close to $\bar{f}_{Q}$. The proof is given in 
\ifshortver
\cite{vector_arxiv}.
\else
Appendix~\ref{subsec:unif_approx_pf}.
\fi
\smallskip
\begin{cor}
\label{cor:unif_approx}Assume $\mathbf{x} \sim \mathrm{Unif}(A)$, where $A\subseteq \mathbb{R}^n$ is convex. If $A\subseteq r\mathbf{G}B_{n}$, where $B_{n}$ is the unit $n$-ball,
and $r>0$, i.e., $A$ is bounded within an ellipsoid $r\mathbf{G}B_{n}$,
then the total variation distance between the distribution of the error  $\mathbf{x}-Q(\mathbf{x})$ and $\bar{f}_{Q}$ is upper-bounded by $n\nu_{n-1} r^{n-1} / \mu(A)$, where $\nu_{n-1}:=\pi^{(n-1)/2}/\Gamma((n+1)/2)$ is the volume of the
unit $(n-1)$-ball.
\end{cor}
\smallskip

As a result, if $\mathbf{x}$ is uniform over the $n$-ball with radius $r$, then the distribution of quantization error will approach $\bar{f}_{Q}$ in total variation as $r\to \infty$.

We now study the use of subtractive dithering to ensure that the error is independent of the input $\mathbf{x}$, and exactly follows $\bar{f}_{Q}$, regardless of whether $\mathbf{x}$ follows a Nyquist-$\mathbf{G}$ distribution. Consider the dither $\mathbf{w}$ with a Nyquist-$\mathbf{G}$ distribution known to both the encoder and the decoder, and is independent of the input $\mathbf{x}$. To quantize $\mathbf{x}$, the encoder encodes $Q(\mathbf{x}+\mathbf{w})$, and the decoder recovers $Q(\mathbf{x}+\mathbf{w})-\mathbf{w}$. By Proposition~\ref{prop:quantize_nyquist}, for any fixed $\mathbf{x}$, the error $\mathbf{x}-(Q(\mathbf{x}+\mathbf{w})-\mathbf{w})$ follows $\bar{f}_{Q}$ exactly since $\mathbf{x}+\mathbf{w}$ follows a Nyquist-$\mathbf{G}$ distribution shifted by $\mathbf{x}$, which is still Nyquist-$\mathbf{G}$. The following corollary follows directly.
% Even if $\mathbf{x}$ does not follow a Nyquist-$\mathbf{G}$ distribution, the distribution of $\mathbf{x}+\mathbf{w}$, where $\mathbf{w}$ is an independent dither with a Nyquist-$\mathbf{G}$ distribution (e.g. uniform over a basic cell), would be a Nyquist-$\mathbf{G}$ distribution \cite[Theorem 2]{kirac1996results}. This allows us to use a shift-periodic quantizer together with subtractive dithering to ensure that the error distribution is exactly $\bar{f}_{Q}$. Consider the dither $\mathbf{w}$ with a Nyquist-$\mathbf{G}$ distribution known to both the encoder and the decoder. To quantize $\mathbf{x}$, the encoder encodes $Q(\mathbf{x}+\mathbf{w})$, and the decoder recovers $Q(\mathbf{x}+\mathbf{w})-\mathbf{w}$. As a corollary of Proposition~\ref{prop:quantize_nyquist}, the error distribution would be $\bar{f}_{Q}$ exactly. The following corollary follows directly from applying Proposition~\ref{prop:quantize_nyquist}
\smallskip
\begin{cor}\label{cor:dither}
Assume the input vector $\mathbf{x}$ follows any
distribution. Consider an independent dither $\mathbf{w}$ with a Nyquist-$\mathbf{G}$ distribution. Then the error $\mathbf{x}-(Q(\mathbf{x}+\mathbf{w})-\mathbf{w})$ is independent of $\mathbf{x}$, and
follows the error distribution $\bar{f}_{Q}$.
\end{cor}
\smallskip

\subsection{Normalized Entropy}

A shift-periodic quantizer can have quantization cells (i.e., preimages
in the form $Q^{-1}(\mathbf{q})$, $\mathbf{q}\in\mathbb{R}^{n}$)
of different shapes, where none of them are basic cells. Nevertheless,
we can find a union of some quantization cells which form a basic
cell, i.e., we find $S \subseteq \mathbb{R}^n$ such that $\bigcup_{\mathbf{q} \in S} Q^{-1}(\mathbf{q}) = Q^{-1}(S)$ is a basic cell. Note that the quantization cell $Q^{-1}(\mathbf{q} + \mathbf{G}\mathbf{v})$ for $\mathbf{v} \in \mathbb{Z}^n$ is just a lattice-shifted version of the cell $Q^{-1}(\mathbf{q})$ (i.e., $Q^{-1}(\mathbf{q} + \mathbf{G}\mathbf{v}) = Q^{-1}(\mathbf{q}) + \mathbf{G}\mathbf{v}$), and they are considered the ``same kind of'' quantization cells. We want to form a basic cell by picking exactly one quantization cell per kind of cells. This can be ensured by considering $Q^{-1}(S)$ where $S$ is a basic cell, so for each kind $\{Q^{-1}(\mathbf{q} + \mathbf{G}\mathbf{v}):\, \mathbf{v} \in \mathbb{Z}^n\}$, there is exactly one $\mathbf{v} \in \mathbb{Z}^n$ such that $\mathbf{q} + \mathbf{G}\mathbf{v} \in S$, and $Q^{-1}(\mathbf{q} + \mathbf{G}\mathbf{v}) \subseteq Q^{-1}(S)$.
This is captured in the definition below.
\smallskip
\begin{defn}
Consider a shift-periodic quantizer $Q$ with generator matrix $\mathbf{G}$.
Given a basic cell $S$ of $\mathbf{G}$, its preimage is $Q^{-1}(S)=\{\mathbf{x}:\,Q(\mathbf{x})\in S\}$,
which is also a basic cell. We call a basic cell a \emph{preimage
basic cell} of $Q$ if it is the preimage of some basic cell.
\end{defn}
\smallskip

%\textcolor{blue}{
% %We define the entropy of a shift-periodic quantizer as follows.
% \begin{defn}
% Consider a shift-periodic quantizer $Q$ with generator matrix $\mathbf{G}$. Let $T$ be a preimage basic cell of $Q$, 
% and $\mathcal{Q} := \{\mathbf{q} = Q(\mathbf{x}):\, \mathbf{x} \in T\}$ be the image of $Q$ (which is finite or countable). 
% For $\mathbf{x} \sim \mathrm{Unif}(T)$, the \emph{entropy} $H(Q(\mathbf{x}))$ of $Q$ is
% \begin{equation} 
% H(Q(\mathbf{x})) := \mathbb{E}\left[- \log \frac{\mu(Q^{-1}(Q(\mathbf{x})))}{|\det \mathbf{G}|} \right] = -\sum_{\mathbf{q}\in \mathcal{Q}} \frac{\mu(Q^{-1}(\mathbf{q}))}{|\det \mathbf{G}|} \log \frac{\mu(Q^{-1}(\mathbf{q}))}{|\det \mathbf{G}|}.
% \end{equation}
% \end{defn}}
% \smallskip

For example, if $Q$ is the lattice quantizer constructed using the basic cell $\tilde{S}$ (i.e., its quantization cells are lattice-translated copies of $\tilde{S}$), then the only preimage basic cells of $Q$ are lattice-translated copies of $\tilde{S}$, i.e., $\tilde{S} + \mathbf{G}\mathbf{v}$ for $\mathbf{v} \in \mathbb{Z}^n$. Refer to Figure \ref{fig:shiftperiodic} for examples of preimage basic cells of a shift-periodic quantizer.

If the input vector $\mathbf{x}$ is random, then the quantized vector $Q(\mathbf{x})$ is random as well. Since the range of $Q$ is countable by definition, we can find the entropy $H(Q(\mathbf{x}))$ of the quantized signal.\footnote{We have $H(Q(\mathbf{x})) = -\sum_{\mathbf{q} \in Q(\mathbb{R}^n)} \mathbf{P}(\mathbf{x} \in Q^{-1}(\mathbf{q})) \log \mathbf{P}(\mathbf{x} \in Q^{-1}(\mathbf{q}))$.}
% {\color{red} Note that for the lattice quantizer with generator matrix $\mathbf{G}$, the entropy $H(Q(\mathbf{x}))$ depends on the input distribution. On the other hand, for the shift-periodic quantizer $Q$ with the generator matrix  $\mathbf{G}$, the entropy $H(Q(\mathbf{x}))$ depends not only on the input distribution but also on the number and shapes of the quantization cells.
% } 
For entropy-constrained quantization~\cite{chou1989entropy}, when using the entropy coding (or variable-rate source coding) after vector quantization, the encoding length is approximately given by the entropy $H(Q(\mathbf{x}))$ of the quantized signal, which depends on the input distribution.  
Therefore, minimizing the entropy amounts to minimizing the average coding length and, consequently, decreasing communication and storage costs.
If $\mathbf{x}$ is uniform over a larger set, the entropy will be larger. 
If $\mathbf{x} \sim \mathrm{Unif}(A)$, we expect $H(Q(\mathbf{x})) = \log \mu(A) \!+\! O(1)$ for large $\mu(A)$. This is because 
\begin{align}
H(Q(\mathbf{x})) & = \mathbb{E}_\mathbf{x}\big[ - \log P_{Q(\mathbf{x})}(\mathbf{x})  \big] \nonumber\\
& = \mathbb{E}_\mathbf{x}\bigg[ - \log \frac{\mu(Q^{-1}(Q(\mathbf{x})) \cap A)}{\mu(A)}  \bigg] \nonumber\\
& = \log \mu(A) + \mathbb{E}_\mathbf{x}\big[ - \log \mu(Q^{-1}(Q(\mathbf{x})) \cap A) \big], \label{eq:HQx_E}
\end{align}
where $P_{Q(\mathbf{x})}$ is the probability mass function of $Q(\mathbf{x})$, and $Q^{-1}(Q(\mathbf{x}))$ is the set of input signals with the same quantization output as $\mathbf{x}$, i.e., the quantization cell containing $\mathbf{x}$. If $A$ is ``large'', we expect $Q^{-1}(Q(\mathbf{x}))$ to be a random quantization cell, where a quantization cell is chosen with probability proportional to its size.
The expectation in~\eqref{eq:HQx_E} can be approximated by the following quantity.
% The $O(1)$ term is captured by the following quantity.
\smallskip
\begin{defn}\label{def:normalized_ent}
The \emph{normalized entropy} of $Q$ is 
\[
\bar{H}(Q):=H(Q(\mathbf{x}))-\log|\det\mathbf{G}|,
\]
where $\mathbf{x}\sim\mathrm{Unif}(T)$,
and $T$ is a preimage basic cell of $Q$.
% is uniform over a preimage
% basic cell $T$ of $Q$.
A straightforward evaluation gives\footnote{Let $T=Q^{-1}(S)$. By \eqref{eq:HQx_E}, we have $\bar{H}(Q) = \mathbb{E}_\mathbf{x}[ - \log \mu(Q^{-1}(Q(\mathbf{x})) \cap T) ]= \mathbb{E}_\mathbf{x}[ - \log \mu(Q^{-1}(Q(\mathbf{x})) ) ]$ since $\mathbf{x} \in T=Q^{-1}(S)$, $Q(\mathbf{x}) \in S$, and $Q^{-1}(Q(\mathbf{x})) \subseteq Q^{-1}(S)=T$.}
\[
\bar{H}(Q) = \mathbb{E}_\mathbf{x}\big[ - \log \mu(Q^{-1}(Q(\mathbf{x}))) \big],
\]
i.e., it is the average of the negative logarithm of the size of a randomly chosen quantization cell, where a quantization cell is chosen with probability proportional to its size.
\end{defn}
\smallskip
Note that $H(Q(\mathbf{x}))$ does not depend on the choice of preimage
basic cell $T$.\footnote{To see this, consider basic cells $S_{1},S_{2}$,
and let $\mathbf{x}\sim\mathrm{Unif}(Q^{-1}(S_{1}))$, and $\mathbf{x}':=\mathbf{x}\,\mathrm{mod}\,Q^{-1}(S_{2})\sim\mathrm{Unif}(Q^{-1}(S_{2}))$,
and $Q(\mathbf{x}')=Q(\mathbf{x})\,\mathrm{mod}\,S_{2}$ (since $Q(\mathbf{x}')-Q(\mathbf{x})$
is a lattice point, and $Q(\mathbf{x}')\in S_{2}$ due to $\mathbf{x}'\in Q^{-1}(S_{2})$),
$Q(\mathbf{x})=Q(\mathbf{x}')\,\mathrm{mod}\,S_{1}$, which implies
$H(Q(\mathbf{x}))=H(Q(\mathbf{x}'))$.} 
Recall that \eqref{eq:HQx_E} suggests that when $\mathbf{x} \sim \mathrm{Unif}(A)$ for a large $A$, $H(Q(\mathbf{x})) \approx \bar{H}(Q) + \log \mu(A)$. We now give a rigorous statement of this fact.
% We then show that, loosely speaking, when $\mathbf{x} \sim \mathrm{Unif}(A)$ for a large $A$, $H(Q(\mathbf{x})) \approx \bar{H}(Q) + \log \mu(A)$. 
The proof is given in 
\ifshortver
\cite{vector_arxiv}.
\else
Appendix~\ref{subsec:ent_large_pf}.
\fi

% We want to define a quantity $\bar{H}(Q)$ is defined in a way that allows us to approximate $H(Q(\mathbf{x}))$, in the sense that $H(Q(\mathbf{x})) \approx \bar{H}(Q) + \log \mu(A)$ when $\mathbf{x} \sim \mathrm{Unif}(A)$ is uniform over a large set. This is made precise in the following proposition.
\smallskip
\begin{prop}
\label{prop:ent_large}
Consider a set $A\subseteq \mathbb{R}^n$ with positive finite volume and measure-zero boundary (i.e., a Jordan measurable set \cite{frink1933jordan}). Let $r>0$, and assume $\mathbf{x} \sim \mathrm{Unif}(rA)$. We have
\[
\lim_{r \to \infty}\big(H(Q(\mathbf{x})) - \log \mu(rA)\big) = \bar{H}(Q).
\]
\end{prop}
\smallskip

We then present a lower bound on the normalized entropy of every shift-periodic quantizer in terms of the differential entropy of the error distribution $h(\bar{f}_{Q})$.
We remark that a similar bound has appeared in \cite{hegazy2022randomized} in the context of randomized scalar quantization. The proof is in 
\ifshortver
\cite{vector_arxiv}.
\else
Appendix~\ref{subsec:lowerbound_pf}.
\fi

\smallskip
\begin{prop}
\label{prop:lowerbound}
For every shift-periodic quantizer $Q:\mathbb{R}^{n}\to\mathbb{R}^{n}$, we have the following lower bound:
\[
\bar{H}(Q)\ge-h(\bar{f}_{Q}).
\]
\end{prop}
\smallskip

In Section~\ref{sec:main}, we will construct shift-periodic quantizers with upper bounds on the normalized entropy. 
In Section~\ref{sec:QuantizerErrorBall}, we will prove upper and lower bounds on the normalized entropy for the case where the error distribution is the uniform distribution over $n$-balls.

\section{Main Results}\label{sec:main}
In this section, we will present constructions of shift-periodic quantizers, and prove upper bounds on their normalized entropies. We first prove a lemma that shows that any set $S\subseteq\mathbb{R}^{n}$
can be dissected into infinitely many pieces to form another set $A\subseteq\mathbb{R}^{n}$
with the same volume, while satisfying an upper bound on the entropy
of the dissection in terms of the volume of the Minkowski difference $S-A$.
We can also terminate this construction early to obtain a quantizer with finitely many quantization cells, but this will result in a difference between the error distribution and $\mathrm{Unif}(A)$, and hence we can only bound the total variation distance between these two distributions.
The proof is given in 
\ifshortver
\cite{vector_arxiv}.
\else
Appendix~\ref{subsec:dissect_pf}.
\fi

\smallskip
\begin{lem}
\label{lem:dissect}Let $A,S\subseteq\mathbb{R}^{n}$ be two bounded sets
with measure-zero boundaries and the same
finite volume. 
\begin{enumerate}
 \item There is a quantizer $F:S\to\mathbb{R}^{n}$ (with possibly infinitely many quantization cells) where, when $\mathbf{u}\sim\mathrm{Unif}(S)$,
 we have $\mathbf{u}-F(\mathbf{u})\sim\mathrm{Unif}(A)$, and 
 \[
 H(F(\mathbf{u}))\le\log\frac{\mu(S-A)}{\mu(A)}+4\;\mathrm{bits}.
 \]
 % \textcolor{red}{In other words, when $\mathbf{x}\sim \mathrm{Unif}(S)$ and $\mu(S)=\mu(A)$, by using $F$, we have $\bar{f}_F=\mathrm{Unif}(A)$ and the entropy of $F$ is bounded as above.}
 \item There is a quantizer $\tilde{F}:S\to\mathbb{R}^{n}$ with at most $k$ quantization cells (i.e., $|\tilde{F}(S)|\le k$) where, when $\mathbf{u}\sim\mathrm{Unif}(S)$,
 we have the following bound on the total variation distance
 \begin{equation} \label{eq:tv}
    \delta_{\mathrm{TV}}\big(\mathbf{u}-\tilde{F}(\mathbf{u}),\, \mathrm{Unif}(A)\big) \le \frac{\mu(S-A)}{\mu(S-A) + k \mu(A)},
 \end{equation}
 and 
 \[
 H(F(\mathbf{u}))\le\log\frac{\mu(S-A)}{\mu(A)}+4\;\mathrm{bits}.
 \]
\end{enumerate}
\end{lem}
\smallskip

We now present our main result.
\smallskip
\begin{thm} \label{thm:quantization}
Let $A\subseteq\mathbb{R}^{n}$ be a bounded set with finite volume and a measure-zero boundary, and $S\subseteq\mathbb{R}^{n}$ be a basic cell of the lattice generated by $\mathbf{G}\in\mathbb{R}^{n\times n}$, with $\mu(S)=\mu(A)$.
\begin{enumerate}
\item There is a shift-periodic quantizer $Q_{1}$ with generator matrix $\mathbf{G}$, with error distribution
$\mathrm{Unif}(A)$, and with normalized entropy upper-bounded by
\begin{equation} \label{eq:q1bound}
\bar{H}(Q_{1})\le\log\frac{\mu(S-A)}{(\mu(A))^{2}}+4\;\mathrm{bits}.
\end{equation}
\item There is a shift-periodic quantizer $Q_{2}$ with generator matrix $\mathbf{G}$, with error distribution
$\mathrm{Unif}(A)$, and with normalized entropy upper-bounded by
\begin{align} \label{eq:q2bound}
\bar{H}(Q_2)\, & \,\leq \,\,
H_{b}\left(\!\frac{\mu(S\backslash A)}{\mu(A)}\!\right)+\frac{\mu(S\backslash A)}{\mu(A)}\left(\!\log\frac{\mu((S\backslash A)-(A\backslash S))}{\mu(S\backslash A)}+4\!\right)\\ \nonumber
&\;\;\; -\log\mu(A)\;\mathrm{bits}. 
\end{align} 
\end{enumerate}
\end{thm}
\smallskip
\begin{IEEEproof}
We prove the first claim. Let $F_1:S\to\mathbb{R}^{n}$ be given by
Lemma \ref{lem:dissect} (first part). The quantizer $Q_{1}:\mathbb{R}^{n}\to\mathbb{R}^{n}$
is defined as
\begin{equation} \label{eq:quantizer1}
Q_{1}(\mathbf{x}):=F_1(\mathbf{x}-\mathbf{G}\mathbf{v})+\mathbf{G}\mathbf{v},
\end{equation}
where $\mathbf{v}\in\mathbb{Z}^{n}$ satisfies $\mathbf{x}\in S+\mathbf{G}\mathbf{v}$.  To compute the error distribution,
since $Q_{1}$ is shift-periodic, it suffices to consider the distribution
of $\mathbf{u}-Q_{1}(\mathbf{u})$ where $\mathbf{u}\sim\mathrm{Unif}(S)$.
% which is the uniform distribution $\mathrm{Unif}(A)$. 
% Conditional on the event $\mathbf{u}\in T_{i}$, we have $\mathbf{u}-Q_{1}(\mathbf{u})=\mathbf{u}-\mathbf{z}_{i}$
% uniform over $T_{i}-\mathbf{z}_{i}$. 
By invoking Lemma~\ref{lem:dissect} (first part),
% Hence, 
when $\mathbf{u}\sim\mathrm{Unif}(S)$,
we have $\mathbf{u}-Q_{1}(\mathbf{u})\sim\mathrm{Unif}(A)$. 
The normalized entropy can be bounded by
\begin{align*}
 H(Q_{1}(\mathbf{u}))-\log|\det\mathbf{G}|
 & =H(Q_{1}(\mathbf{u}))-\log\mu(A)\\
 & \le\log\frac{\mu(S-A)}{(\mu(A))^{2}}+4\;\mathrm{bits}.
\end{align*}

We then prove the second claim. Let $\hat{A}:=A\backslash S$, $\hat{S}:=S\backslash A$.
Applying Lemma \ref{lem:dissect} (first part) on $\hat{A},\hat{S}$, we obtain
$F_2:\hat{S}\to\mathbb{R}^{n}$. The quantizer $Q_{2}:\mathbb{R}^{n}\to\mathbb{R}^{n}$
is defined as
\begin{equation} \label{eq:quantizer2}
Q_{2}(\mathbf{x}):=\begin{cases}
\mathbf{G}\mathbf{v} & \mathrm{if}\;\mathbf{x}-\mathbf{G}\mathbf{v}\in A\\
F_2(\mathbf{x}-\mathbf{G}\mathbf{v})+\mathbf{G}\mathbf{v} & \mathrm{if}\;\mathbf{x}-\mathbf{G}\mathbf{v}\notin A,
\end{cases}
\end{equation}
where $\mathbf{v}\in\mathbb{Z}^{n}$ satisfies $\mathbf{x}\in S+\mathbf{G}\mathbf{v}$.
It can be checked that $\mathbf{u}-Q_{2}(\mathbf{u})\sim\mathrm{Unif}(A)$
when $\mathbf{u}\sim\mathrm{Unif}(S)$. We have
\begin{align*}
 & H(Q_{2}(\mathbf{u}))-\log|\det\mathbf{G}|\\
 & =H_{b}\left(\mathbb{P}\left(\mathbf{u}\in A\right)\right)+\mathbb{P}\left(\mathbf{u}\notin A\right)H(Q(\mathbf{u})\,|\,\mathbf{u}\notin A)-\log\mu(A)\\
 & \le H_{b}\left(\!\frac{\mu(S\backslash A)}{\mu(A)}\!\right)+\frac{\mu(S\backslash A)}{\mu(A)}\!\left(\!\log\frac{\mu(\hat{S}-\hat{A})}{\mu(\hat{A})}+4\!\right)\! -\log\mu(A)\\
 & =H_{b}\left(\frac{\mu(S\backslash A)}{\mu(A)}\right)+\frac{\mu(S\backslash A)}{\mu(A)}\left(\log\frac{\mu((S\backslash A)-(A\backslash S))}{\mu(S\backslash A)}+4\right)\\
 &\;\;\; -\log\mu(A)\;\mathrm{bits}.
\end{align*}
An illustration of $Q_2(\mathbf{x})$ when $A$ is the unit disk and $S$ is a regular hexagon is given in Figure~\ref{fig::constructionF}.
\end{IEEEproof}

\medskip

Notice that when $A \cap S = \emptyset$, then the quantizers $Q_{1}$ and $Q_{2}$ are equivalent, and \eqref{eq:q1bound} coincides with \eqref{eq:q2bound}. The advantage of \eqref{eq:q2bound} over \eqref{eq:q1bound} is that if $A \cap S$ is large, then the first case in \eqref{eq:quantizer2} occurs with large probability, which results in a small normalized entropy. This agrees with the intuition that if $A$ is close to a basic cell, then the normalized entropy is small. 
When $A=S$, it is easy to see that $\bar{H}(Q_2) \leq -\log \mu(S)$ (by (\ref{eq:q2bound}) in Theorem~\ref{thm:quantization}) 
and $\bar{H}(Q_2) \geq -\log \mu(S)$  (by Proposition~\ref{prop:lowerbound}), implying $\bar{H}(Q_2)=-\log \mu(S)$,
which is the normalized entropy of the conventional lattice quantizer with error uniform over the basic cell $S$.
% This implies that the entropy $H(Q(\mathbf{x})) = 0$ when $\mathbf{x}$ is uniform over a preimage basic cell, confirming the entropy of the conventional lattice quantizer when the input distribution is uniform over the basic cell $S$.

The quantizers constructed in Theorem \ref{thm:quantization} have infinitely many quantization cells. For any basic cell $S$, the size of the set $Q(\mathbb{R}^n)\cap S$ (which is the ``number of quantization cells per basic cell'') is infinite. If a simpler quantizer with only finitely many quantization cells per basic cell is desired, we can 
% terminate the construction in Lemma~\ref{lem:dissect} early 
invoke the second part in Lemma~\ref{lem:dissect}
to obtain the following results. 

\medskip

\begin{thm} \label{thm:quantization_tv}
Let $A\subseteq\mathbb{R}^{n}$ be a bounded set with finite volume and a measure-zero boundary,  $S\subseteq\mathbb{R}^{n}$ be a basic cell of the lattice generated by $\mathbf{G}\in\mathbb{R}^{n\times n}$, with $\mu(S)=\mu(A)$.
\begin{enumerate}
\item For $k \in \mathbb{Z}$, $k\ge 1$, there is a shift-periodic quantizer $\tilde{Q}_{1}$ with generator matrix $\mathbf{G}$, with a normalized entropy upper-bounded by \eqref{eq:q1bound}, $|\tilde{Q}_{1}(\mathbb{R}^n)\cap \tilde{S}| \le k$ for any basic cell $\tilde{S}$, and an error distribution close to $\mathrm{Unif}(A)$ in the sense that
\begin{align}
\delta_{\mathrm{TV}}(\bar{f}_{\tilde{Q}_1},\mathrm{Unif}(A)) &\le \frac{\mu(S-A)}{\mu(S-A) + k \mu(A)}. \label{eq:q1tbound}
\end{align}
\item For $k \in \mathbb{Z}$, $k\ge 2$, there is a shift-periodic quantizer $\tilde{Q}_{2}$ with generator matrix $\mathbf{G}$, with a normalized entropy upper-bounded by \eqref{eq:q2bound}, $|\tilde{Q}_{2}(\mathbb{R}^n)\cap \tilde{S}| \le k$ for any basic cell $\tilde{S}$, and an error distribution close to $\mathrm{Unif}(A)$ in the sense that
\begin{align}
\delta_{\mathrm{TV}}(\bar{f}_{\tilde{Q}_2},\mathrm{Unif}(A)) &\le  \frac{\mu(S\backslash A-A\backslash S)}{\mu(S\backslash A-A\backslash S) + (k-1) \mu(A\backslash S)}. \label{eq:q2tbound}
\end{align}
\end{enumerate}
\end{thm}
\begin{IEEEproof}
We prove the first claim. Let $\tilde{F}_1:S\to\mathbb{R}^{n}$ be the quantizer with $k$ quantization cells given by
Lemma \ref{lem:dissect} (second part). The quantizer $\tilde{Q}_{1}:\mathbb{R}^{n}\to\mathbb{R}^{n}$
is defined as
\begin{equation} 
\tilde{Q}_{1}(\mathbf{x}):=\tilde{F}_1(\mathbf{x}-\mathbf{G}\mathbf{v})+\mathbf{G}\mathbf{v},
\end{equation}
where $\mathbf{v}\in\mathbb{Z}^{n}$ satisfies $\mathbf{x}\in S+\mathbf{G}\mathbf{v}$.  
Since $\tilde{F}_1$ partitions $S$ into at most $k$ quantization cells, $\tilde{Q}_{1}$ has at most $k$ quantization cells per basic cell.
To compute the error distribution,
since $\tilde{Q}_{1}$ is shift-periodic, it suffices to consider the distribution
of $\mathbf{u}-\tilde{Q}_{1}(\mathbf{u})$ where $\mathbf{u}\sim\mathrm{Unif}(S)$, i.e., $\bar{f}_{\tilde{Q}_1}$.
% By (\ref{eq:tv}), we have 
% \[\delta_{\mathrm{TV}}(\bar{f}_{\tilde{Q}_1},\mathrm{Unif}(A)) \le \frac{\mu(S-A)}{\mu(S-A) + k \mu(A)}.\]
We obtain \eqref{eq:q1tbound} as a consequence of \eqref{eq:tv}.
We have the same bound as \eqref{eq:q1bound} by Lemma \ref{lem:dissect} (second part).

We then prove the second claim. Let $\hat{A}:=A\backslash S$, $\hat{S}:=S\backslash A$.
Applying Lemma \ref{lem:dissect} (second part) on $\hat{A},\hat{S}$, we obtain the quantizer 
$\tilde{F}_2:\hat{S}\to\mathbb{R}^{n}$ with $k-1$ quantization cells. The quantizer $\tilde{Q}_{2}:\mathbb{R}^{n}\to\mathbb{R}^{n}$
is defined as
\begin{equation} 
\tilde{Q}_{2}(\mathbf{x}):=\begin{cases}
\mathbf{G}\mathbf{v} & \mathrm{if}\;\mathbf{x}-\mathbf{G}\mathbf{v}\in A\\
\tilde{F}_2(\mathbf{x}-\mathbf{G}\mathbf{v})+\mathbf{G}\mathbf{v} & \mathrm{if}\;\mathbf{x}-\mathbf{G}\mathbf{v}\notin A,
\end{cases}\label{eq:q2t_def}
\end{equation}
where $\mathbf{v}\in\mathbb{Z}^{n}$ satisfies $\mathbf{x}\in S+\mathbf{G}\mathbf{v}$.
Since $\tilde{F}_2$ partitions $S$ into at most $k-1$ quantization cells, $\tilde{Q}_{2}$ has at most $k$ quantization cells per basic cell (one of them comes from the first case in \eqref{eq:q2t_def}).
To compute the error distribution,
since $\tilde{Q}_{2}$ is shift-periodic, it suffices to consider the distribution
of $\mathbf{u}-\tilde{Q}_{2}(\mathbf{u})$ where $\mathbf{u}\sim\mathrm{Unif}(S)$, i.e., $\bar{f}_{\tilde{Q}_2}$.
Substituting $S$ by $S \backslash A$ and $A$ by $A\backslash S$ in (\ref{eq:tv}), we can obtain \eqref{eq:q2tbound}.
% \[\delta_{\mathrm{TV}}(\bar{f}_{\tilde{Q}_2},\mathrm{Unif}(A)) \le \frac{\mu(S\backslash A-A\backslash S)}{\mu(S\backslash A-A\backslash S) + (k-1) \mu(A\backslash S)}.\]
We have the same bound by \eqref{eq:q2bound} by Lemma \ref{lem:dissect} (second part).
\end{IEEEproof}

\medskip

\section{Quantizers with Error Uniform over $n$-Balls}\label{sec:QuantizerErrorBall}

In this section, we study the case where the error distribution is the uniform distribution over the $n$-ball. Applying Theorem~\ref{thm:quantization} on $A=B_n$, there is a shift-periodic quantizer with error distribution
uniform over the $n$-ball, with the following bound on the normalized
entropy. The proof is given in 
\ifshortver
\cite{vector_arxiv}.
\else
Appendix~\ref{subsec:cor_ball_pf}.
\fi

\medskip
\begin{cor}
\label{cor:ball}Fix $n\ge1$. There is a shift-periodic quantizer $Q$
with error distribution that is the uniform distribution over the
unit $n$-ball $B_{n}$, and with normalized entropy upper-bounded
by
\begin{align*}
\bar{H}(Q) &\le n\log\left(\sqrt{\frac{\pi e}{2}}+1\right)-\log\mu(B_{n})+4 \;\mathrm{bits} \\
&\le 1.617 n-\log\mu(B_{n})+4 \;\mathrm{bits}. 
% 1.6165297547346775716126757367963213619213416967226352003059369004...
\end{align*}
\end{cor}
\medskip

We now prove a converse result, i.e., a lower bound on $\bar{H}(Q)$ that a shift-periodic quantizer with error uniform over $B_n$ must satisfy. Invoking Proposition~\ref{prop:lowerbound}, we can lower bound $\bar{H}(Q)$ in terms of the size of $B_n$:
\begin{align}
    \bar{H}(Q) \geq  - \log \mu(B_{n}).\label{eq:hq_lb_1}
\end{align}
Note that a lattice quantizer, with error distribution which is uniform over the basic cell $S$, has a normalized entropy $-\log \mu(S)$. Therefore, we could attain the lower bound in \eqref{eq:hq_lb_1} if there was a lattice with a basic cell that is the ball $B_n$. Unfortunately, this is impossible since $B_n$ cannot tile $\mathbb{R}^n$. There must be a gap in \eqref{eq:hq_lb_1} incurred by sphere packing, which cannot attain a $100\%$ density.

Therefore, we will prove a lower bound tighter than \eqref{eq:hq_lb_1} in terms of the sphere packing density. 
The main idea is that, if we have a quantizer with $\bar{H}(Q) \approx -\log \mu(B_n)$, then there must be quantization cells that are close to $B_n$. We can then fit slightly smaller balls in these quantization cells, so that these smaller balls in different quantization cells do not overlap. These smaller balls will form a sphere packing over $\mathbb{R}^n$, and hence cannot have a density that exceeds the optimal sphere packing density.
We state the result below, and defer the proof to Appendix~\ref{subsec:thm_ball_pf}.
\medskip
\begin{thm} \label{thm:ub_on_ball}
Consider any shift-periodic quantizer $Q:\mathbb{R}^{n} \to \mathbb{R}^{n}$ with 
error distribution $\mathrm{Unif}(B_n)$. We must have
\[
    \bar{H}(Q) \geq \left(1- \frac{4^{n/(n+1)}\eta_n}{\left(4^{1/(n+1)}- (1-\eta_{n}^{1/n})\right)^n} \right) \Bigg(\frac{(1-\eta_{n}^{1/n})^{(n+1)/2}}{2\sqrt{2\pi n}}\Bigg) \log e - \log \mu(B_{n}),
\]
where $\eta_n$ is the optimal density of sphere packing over $\mathbb{R}^n$.
\end{thm}
\medskip

% \textcolor{red}{For a concrete bound, we can substitute the bound $\eta_n \le ?????$ [??????] to obtain
% \[
% ?????????????
% \]
% }

The bound in Corollary~\ref{cor:ball} is quite loose. In the following two examples, we will study constructions of quantizers with error distribution
uniform over the 2D disk and the 3D ball with better bounds.

\begin{figure}
    \centering
    \includegraphics[width=250pt]{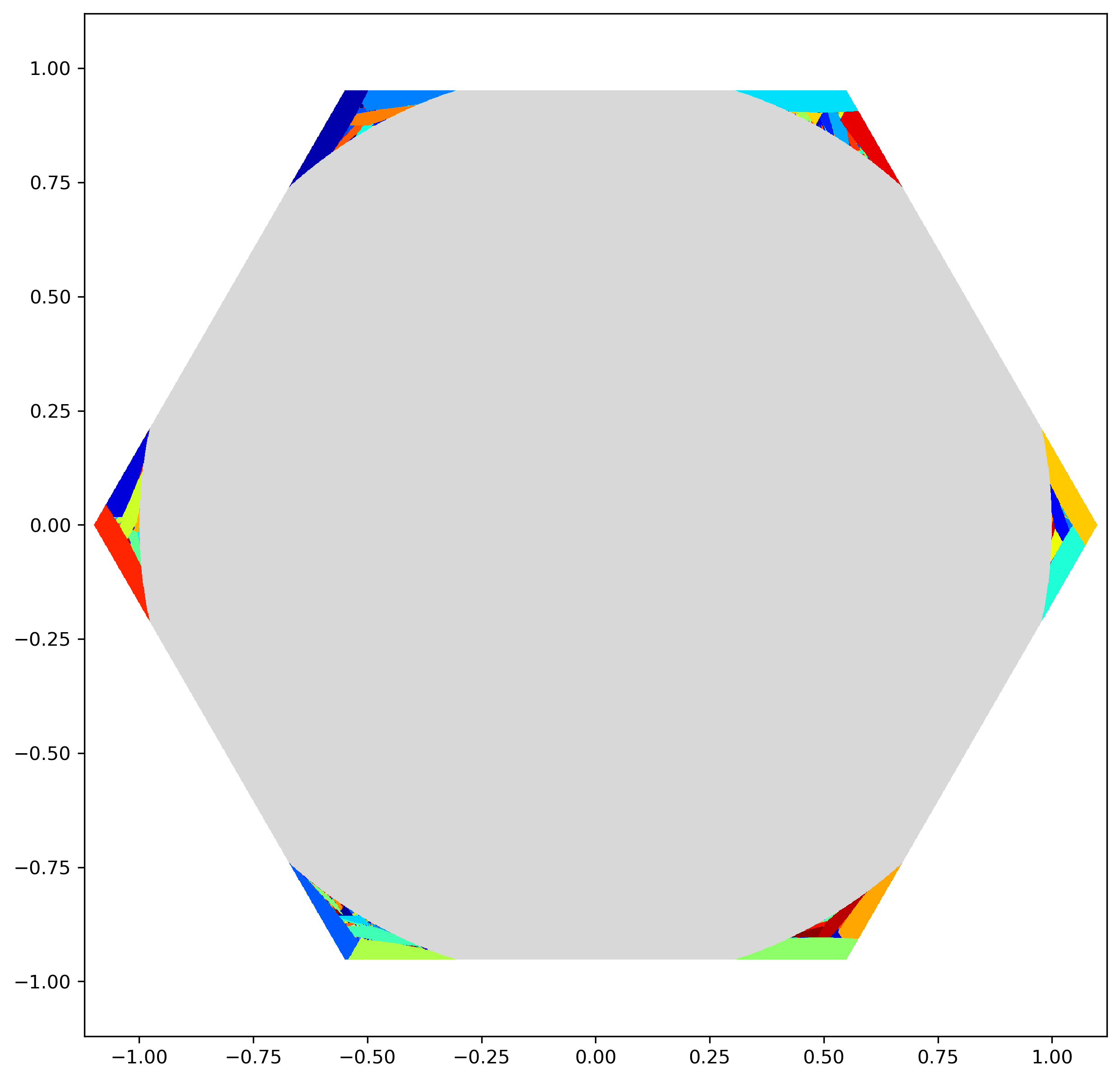}
    \includegraphics[width=250pt]{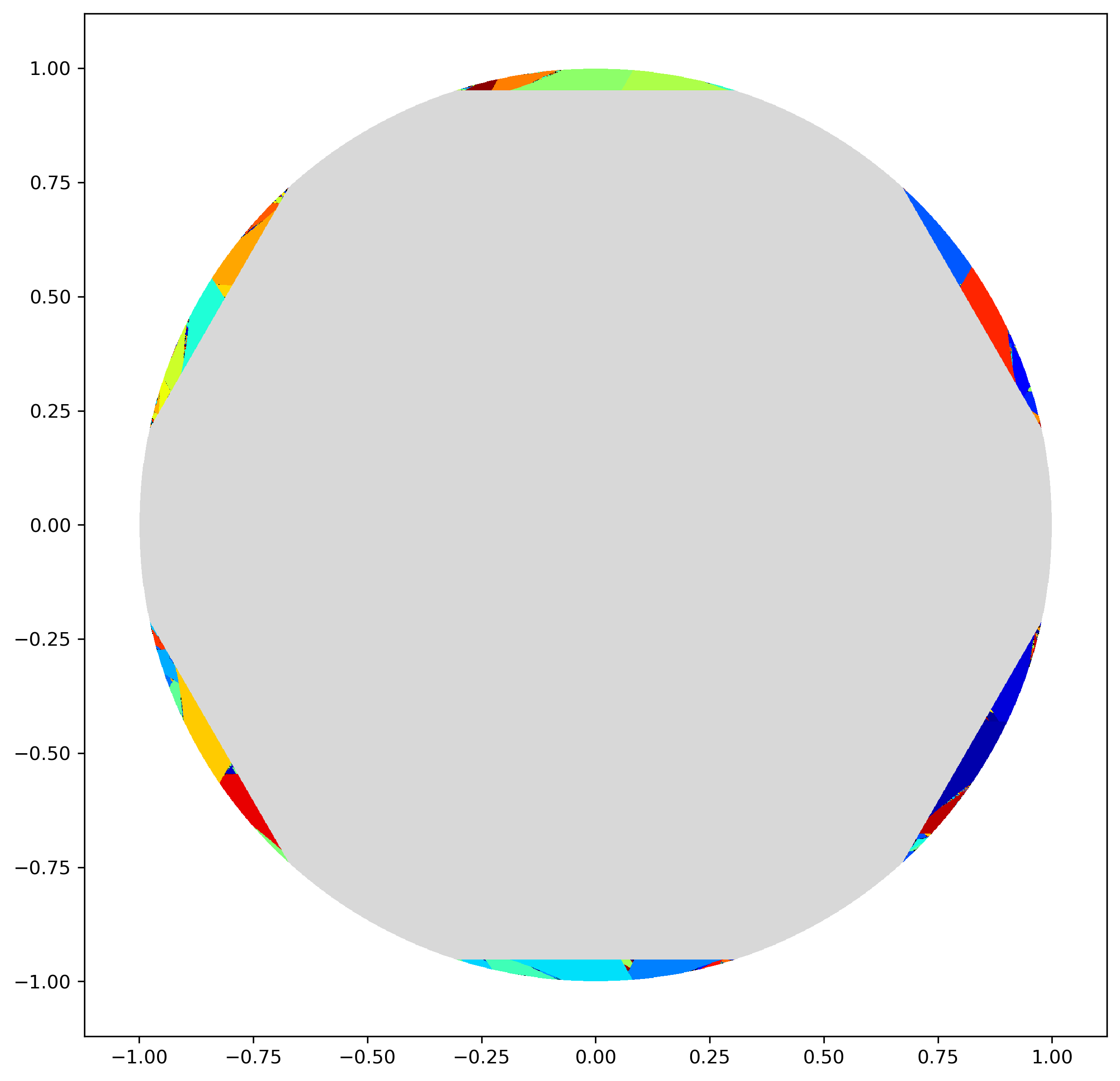}
    \caption{The quantizer constructed in Lemma~\ref{lem:dissect} and Theorem~\ref{thm:quantization} (second part) based on the hexagonal lattice, with an error distribution uniform over the unit disk. The left figure shows a basic cell of the lattice, partitioned into infinitely many quantization cells, each with a different color. The right figure shows how these quantization cells can be translated to form the unit disk. For example, if the input vector lies in the red region on the left tip of the hexagon, it would be quantized to an appropriate vector such that the quantization error will lie in the red region of the same shape on the top-right edge of the disk.
    \label{fig::constructionF}
    }
\end{figure}

\smallskip
\begin{example}
\label{exa:disk}
We now study a shift-periodic quantizer over $\mathbb{R}^2$ with an error distribution that is uniform over the unit disk. Consider the unit disk $B_2 \subseteq \mathbb{R}^2$ centered at the origin, and a basic cell $S$ given by the fundamental Voronoi region induced by the hexagonal lattice (which is a regular hexagon). 
We choose $S$ to be a hexagon (instead of a square or other possible basic cells) since it is closest to a disk, and hence gives a smaller $\mu(S\backslash B_2)$.
Theorem~\ref{thm:quantization} (second part) gives a shift-periodic quantizer
\[
\bar{H}(Q) \le -1.01666 \;\mathrm{bits}.
\]
The calculations are deferred to 
\ifshortver
\cite{vector_arxiv}.
\else
Appendix~\ref{subsec:disk_pf}.
\fi
On the other hand, Proposition~\ref{prop:lowerbound} shows that all shift-periodic quantizers with error distribution $\mathrm{Unif}(B_2)$ satisfies $\bar{H}(Q) \ge -\log \pi \approx -1.65150$, %-1.65149612947
while Theorem~\ref{thm:ub_on_ball} gives a slightly tighter lower bound $\bar{H}(Q) \ge -1.65142$.
%calculated value is -1.6514197289528356
The quantizer constructed in Lemma~\ref{lem:dissect} and Theorem~\ref{thm:quantization} (second part) is given in Figure~\ref{fig::constructionF}. Finding the optimal $\bar{H}(Q)$, and whether the hexagon construction is optimal, are left for future studies.
\end{example}

\smallskip

\begin{example}
\label{exa:ball}
We then study a shift-periodic quantizer over $\mathbb{R}^3$ with an error distribution that is uniform over the unit ball.
Consider the unit ball $B_3$ centered at the origin, and a basic cell $S_{1}$ given by the fundamental Voronoi region induced by the face-centered cubic lattice (which is a rhombic dodecahedron). Theorem~\ref{thm:quantization} (second part) gives a shift-periodic quantizer
\[
\bar{H}(Q) \le -0.77892 \;\mathrm{bits}.
\]
Alternatively, if we consider the fundamental Voronoi region $S_{2}$  induced by the body-centered cubic lattice (which is a truncated octahedron) instead, Theorem~\ref{thm:quantization} (second part) gives a shift-periodic quantizer
\[
\bar{H}(Q) \le -0.74221 \;\mathrm{bits},
\]
which is slightly worse.
The calculations are deferred to 
\ifshortver
\cite{vector_arxiv}.
\else
Appendix~\ref{subsec:ball_pf}.
\fi 
Proposition~\ref{prop:lowerbound} shows that all shift-periodic quantizers with error distribution $\mathrm{Unif}(B_3)$ satisfies $\bar{H}(Q) \ge -\log((4/3) \pi) \approx -2.06653$, %the calculated value is -2.06653362875
while Theorem~\ref{thm:ub_on_ball} gives a slightly tighter lower bound $\bar{H}(Q) \ge -2.06641$
% -2.0664021537921036
. 
Finding the optimal $\bar{H}(Q)$, and whether the face-centered cubic lattice or the body-centered cubic lattice is optimal, are left for future studies.
\end{example}
\smallskip

\section{Nonuniform Error Distribution}\label{sec:nonunif}

% In this section, we present a layered construction, called \emph{layered shift-periodic quantizer}, by using the slicing technique studied in \cite{hegazy2022randomized}. 
In the previous sections, we have studied shift-periodic quantizers where the error distribution approaches a uniform distribution when the signal is uniform over an increasingly large set (Corollary \ref{cor:unif_approx}), and is exactly a uniform distribution when dithering is applied (Corollary \ref{cor:dither}). We now generalize this construction to allow the error to follow any continuous (uniform/nonuniform) distribution when dithering is applied, using the layered quantization technique in \cite{hegazy2022randomized}.
The construction is based on the observation that any continuous distribution (with a probability density function $f_{\mathbf{z}}(\mathbf{z})$) of an $n$-dimensional random vector $\mathbf{z} \in \mathbb{R}^n$ can be expressed as a mixture of uniform distributions $\mathrm{Unif}(A_r)$, where 
\[
A_r := \big\{ \mathbf{x} \in \mathbb{R}^n:\, f_{\mathbf{z}}(\mathbf{x}) \ge r \big\}
\]
is the superlevel set of $f_{\mathbf{z}}$ at $r \in (0,\, \sup_{\mathbf{z}}f_{\mathbf{z}}(\mathbf{z}))$.  To show this, note that if $(\mathbf{z},R) \sim \mathrm{Unif}(\{ (\mathbf{x},r) \in \mathbb{R}^{n+1}:\, 0 \le r \le f_{\mathbf{z}}(\mathbf{x}) \})$, then $\mathbf{z} \sim f_{\mathbf{z}}$ (this fact has been referred as the \emph{fundamental theorem of simulation} \cite{robert2004monte}). Also, the conditional distribution of $\mathbf{z}$ given $R=r$ is $\mathrm{Unif}(A_r)$, and the probability density function of $R$ is $f_R(r) := \mu(A_r)$ for $r \in (0,\, \sup_{\mathbf{z}}f_{\mathbf{z}}(\mathbf{z}))$. Therefore, if we generate $R \sim f_R$, and generate $\mathbf{z}$ from $\mathrm{Unif}(A_R)$, then the overall distribution of $\mathbf{z}$ is $f_\mathbf{z}$.

Therefore, we can create an ensemble of shift-periodic quantizers with an ``average error distribution'' $f_\mathbf{z}$. The definition of such an ensemble is given below.

% $(Q_r)_{r \in (0,\, \sup_{\mathbf{z}}f_{\mathbf{z}}(\mathbf{z}))}$, where $Q_r$ is a shift-periodic quantizer with error distribution $\bar{f}_{Q_r}$ being the uniform distribution $\mathrm{Unif}(D_r)$. For example, we can construct $Q_r$ using Theorem~\ref{thm:quantization}. 

% We now give a layered construction by using Theorem~\ref{thm:quantization} (first part).
\smallskip
\begin{defn}
Given a probability density function $f_{\mathbf{z}}(\mathbf{z})$ over $\mathbb{R}^n$, 
we call $(Q_r)_{r \in (0,\, \sup_{\mathbf{z}}f_{\mathbf{z}}(\mathbf{z}))}$ a \emph{layered shift-periodic quantizer ensemble} for the error distribution $f_{\mathbf{z}}(\mathbf{z})$ if $Q_r$ is a shift-periodic quantizer with error distribution $\bar{f}_{Q_r}$ being the uniform distribution $\mathrm{Unif}(A_r)$ for $r \in (0,\, \sup_{\mathbf{z}}f_{\mathbf{z}}(\mathbf{z}))$ (e.g., we can construct $Q_r$ using Theorem~\ref{thm:quantization}).
% the \emph{layered shift-periodic quantizer with dithering}, denoted as  $Q_R(\mathbf{x})$, is defined using the random state $R \sim f_R$, where $f_R(r):= \mu(A_r)$ for $r>0$, and the shift-periodic quantizer $Q_{1}:\mathbb{R}^n \rightarrow \mathbb{R}^n$.
% We take $A_r =A_r$, find a corresponding full-rank generator matrix $\mathbf{G}_r \in \mathbb{R}^n$ with basic cell $S_r$, generate a dither $\mathbf{w}_r \sim \mathrm{Unif}(A_r)$, and apply $Q$ to $A_r$ and $S_r$ with 
% \begin{equation} \label{eq:layeredQuantizer}
% Q_{r}(\mathbf{x})
% :=Q(\mathbf{x}+\mathbf{w}_r)-\mathbf{w}_{r} 
% \end{equation}
% where $\mathbf{v}\in \mathbb{Z}^n$ satisfies $\mathbf{x} \in S_r+\mathbf{G}_r\mathbf{v}$.
% In other words, the layered shift-periodic quantizer $Q_R(\mathbf{x})$ is a collection of shift-periodic quantizers with dithering $Q_r(\mathbf{x}):\mathbb{R}^n \rightarrow \mathbb{R}^n$ in (\ref{eq:layeredQuantizer}), parametrized by $r$.
% , each with a full-rank generator matrix $\mathbf{G}_r \in \mathbb{R}^n$ and a basic cell $S_r$ parameterized by $r$.
\end{defn}
\smallskip

If we choose a random quantizer $Q_R$ in a layered shift-periodic quantizer ensemble where $R \sim f_R$ is randomly generated, then the average error distribution will be $f_{\mathbf{z}}$. To make the error distribution exact, we may apply dithering as in Corollary \ref{cor:dither}. 
% Given a shift-periodic quantizer ensemble, we can perform a dithered quantization as follows. 
First, generate a random state $R \sim f_R$ and a random dither $\mathbf{w}$ uniform over a basic cell of the lattice $\Lambda(\mathbf{G}_R)$, where $\mathbf{G}_r$  is the generator matrix of $Q_r$. The pair $(R,\mathbf{w})$ is a common randomness known to both the encoder and the decoder. To quantize $\mathbf{x}$, the encoder produces $Q_R(\mathbf{x}+\mathbf{w})$, and the decoder recovers $\mathbf{y}=Q_R(\mathbf{x}+\mathbf{w})-\mathbf{w}$. Since the error distribution for a fixed $R=r$ is uniform over $A_r$ (Corollary \ref{cor:dither}), the overall error distribution is the mixture of $\mathrm{Unif}(A_r)$, which is $f_{\mathbf{z}}$.

In particular, we are interested in having an error distributed according to the multivariate Gaussian distribution, multivariate Laplace distribution~\cite{andres2013geo}, or another elliptical distribution \cite{reimherr2019elliptical}, which is useful for differential privacy \cite{Dwork06DP,Dwork14Book,duchi2013local} and machine learning \cite{flamich2020compressing}. The superlevel set $A_r$ of the probability density function of a unimodal elliptical distribution\footnote{Here, a unimodal elliptical distribution is a distribution where the density $f_{\mathbf{z}}$ has superlevel sets $A_r = g(r)\mathbf{M} B_n + \mathbf{c}$ for all $r>0$ with $A_r \neq \emptyset$, where $g:(0,\infty) \to [0, \infty)$, $\mathbf{M}\in \mathbb{R}^{n\times n}$ is full rank, and $\mathbf{c} \in \mathbb{R}^n$.} is always an ellipsoid. 
Recall that we have studied the construction of shift-periodic quantizers with error uniform over a ball in Corollary \ref{cor:ball}. Applying a suitable linear transformation, we can construct a shift-periodic quantizer $Q_r$ with error distribution $\mathrm{Unif}(A_r)$, giving a layered shift-periodic quantizer ensemble with an overall error following the desired unimodal elliptical distribution. We leave the analyses on the performance of this construction to future studies.

We remark that ensembles of quantizers have also been studied in distribution preserving quantization \cite{li2010distribution}, where the goal is to control the distribution of the output $\mathbf{y}$. In this paper, our goal is not to control the distribution of $\mathbf{y}$, but rather the distribution of the error $\mathbf{y}-\mathbf{x}$.

\smallskip

\section{Conclusion and Discussions}
In this paper, we studied quantizers where the error distribution can be uniform over any given set, and proved several bounds on their normalized entropies.

Since any continuous nonuniform distribution can be expressed as a mixture of uniform distributions, the quantizers in this paper can be applied in a randomized manner (similar to \cite{hegazy2022randomized}) to allow the error distribution to be any continuous (uniform/nonuniform) distribution. 
We have described the layered shift-periodic quantizer ensemble with dithering for any continuous (uniform/nonuniform) error distribution, and discussed how to apply such a quantizer ensemble to ensure that the error follows a Gaussian distribution or a Laplace distribution.
Bounds on the entropy given by such construction is left for future studies.

The optimal normalized entropy of shift-periodic quantizers with error distribution uniform over the $n$-ball is still unknown. For $n=2$, we conjecture that the optimal quantizer is based on the hexagonal lattice as in Example~\ref{exa:disk}. For $n=3$, we conjecture that the optimal quantizer is based on the face-centered cubic lattice as in Example~\ref{exa:ball}.

\smallskip

\section{Acknowledgement}

% The work of Cheuk Ting Li was supported in part by the Hong Kong Research Grant Council Grant ECS No. CUHK 24205621.
The work described in this paper was partially supported by an
ECS grant from the Research Grants Council of the Hong Kong Special Administrative
Region, China [Project No.: CUHK 24205621].
The authors would like to thank Yanxiao Liu for the insightful discussions.
The authors would like to thank the anonymous reviewers of ISIT2023 for their valuable comments on the conference version~\cite{vectquant_isit}.

\bibliographystyle{IEEEtran}
\bibliography{ref}

\begin{thebibliography}{10}
\providecommand{\url}[1]{#1}
\csname url@samestyle\endcsname
\providecommand{\newblock}{\relax}
\providecommand{\bibinfo}[2]{#2}
\providecommand{\BIBentrySTDinterwordspacing}{\spaceskip=0pt\relax}
\providecommand{\BIBentryALTinterwordstretchfactor}{4}
\providecommand{\BIBentryALTinterwordspacing}{\spaceskip=\fontdimen2\font plus
\BIBentryALTinterwordstretchfactor\fontdimen3\font minus
  \fontdimen4\font\relax}
\providecommand{\BIBforeignlanguage}[2]{{%
\expandafter\ifx\csname l@#1\endcsname\relax
\typeout{** WARNING: IEEEtran.bst: No hyphenation pattern has been}%
\typeout{** loaded for the language `#1'. Using the pattern for}%
\typeout{** the default language instead.}%
\else
\language=\csname l@#1\endcsname
\fi
#2}}
\providecommand{\BIBdecl}{\relax}
\BIBdecl

\bibitem{conway2013sphere}
J.~H. Conway and N.~J.~A. Sloane, \emph{Sphere packings, lattices and
  groups}.\hskip 1em plus 0.5em minus 0.4em\relax Springer Science \& Business
  Media, 2013, vol. 290.

\bibitem{zamir2014}
R.~Zamir, B.~Nazer, Y.~Kochman, and I.~Bistritz, \emph{Lattice Coding for
  Signals and Networks: A Structured Coding Approach to Quantization,
  Modulation and Multiuser Information Theory}.\hskip 1em plus 0.5em minus
  0.4em\relax Cambridge University Press, 2014.

\bibitem{gersho1979}
A.~Gersho, ``Asymptotically optimal block quantization,'' \emph{IEEE
  Transactions on Information Theory}, vol.~25, no.~4, pp. 373--380, 1979.

\bibitem{conway1982}
J.~Conway and N.~Sloane, ``Voronoi regions of lattices, second moments of
  polytopes, and quantization,'' \emph{IEEE Transactions on Information
  Theory}, vol.~28, no.~2, pp. 211--226, 1982.

\bibitem{Toth1959SurLR}
L.~F. T{\'o}th, ``Sur la repr{\'e}sentation d'une population infinie par un
  nombre fini d'{\'e}l{\'e}ments,'' \emph{Acta Mathematica Hungarica}, vol.~10,
  pp. 299--304, 1959.

\bibitem{barnes1983}
E.~S. Barnes and N.~J.~A. Sloane, ``The optimal lattice quantizer in three
  dimensions,'' \emph{SIAM Journal on Algebraic Discrete Methods}, vol.~4,
  no.~1, pp. 30--41, 1983.

\bibitem{gray1998quantization}
R.~M. Gray and D.~L. Neuhoff, ``Quantization,'' \emph{IEEE transactions on
  information theory}, vol.~44, no.~6, pp. 2325--2383, 1998.

\bibitem{bennett1948spectra}
W.~R. Bennett, ``Spectra of quantized signals,'' \emph{The Bell System
  Technical Journal}, vol.~27, no.~3, pp. 446--472, 1948.

\bibitem{gariby2008general}
T.~Gariby and U.~Erez, ``On general lattice quantization noise,'' in \emph{2008
  IEEE International Symposium on Information Theory}.\hskip 1em plus 0.5em
  minus 0.4em\relax IEEE, 2008, pp. 2717--2721.

\bibitem{roberts1962}
L.~Roberts, ``Picture coding using pseudo-random noise,'' \emph{IRE
  Transactions on Information Theory}, vol.~8, no.~2, pp. 145--154, 1962.

\bibitem{schuchman1964dither}
L.~Schuchman, ``Dither signals and their effect on quantization noise,''
  \emph{IEEE Transactions on Communication Technology}, vol.~12, no.~4, pp.
  162--165, 1964.

\bibitem{Limb1969}
J.~O. Limb, ``Design of dither waveforms for quantized visual signals,''
  \emph{The Bell System Technical Journal}, vol.~48, no.~7, pp. 2555--2582,
  1969.

\bibitem{Jaynant1972}
N.~S. Jayant and L.~R. Rabiner, ``The application of dither to the quantization
  of speech signals,'' \emph{The Bell System Technical Journal}, vol.~51,
  no.~6, pp. 1293--1304, 1972.

\bibitem{sripad1977necessary}
A.~Sripad and D.~Snyder, ``A necessary and sufficient condition for
  quantization errors to be uniform and white,'' \emph{IEEE Transactions on
  Acoustics, Speech, and Signal Processing}, vol.~25, no.~5, pp. 442--448,
  1977.

\bibitem{Zamir1996Noise}
R.~Zamir and M.~Feder, ``On lattice quantization noise,'' \emph{Information
  Theory, IEEE Transactions on}, vol.~42, pp. 1152 -- 1159, 08 1996.

\bibitem{Zador1982asymptotic}
P.~Zador, ``Asymptotic quantization error of continuous signals and the
  quantization dimension,'' \emph{IEEE Transactions on Information Theory},
  vol.~28, no.~2, pp. 139--149, 1982.

\bibitem{Gray2002onZador}
R.~Gray, T.~Linder, and J.~Li, ``A lagrangian formulation of zador's
  entropy-constrained quantization theorem,'' \emph{IEEE Transactions on
  Information Theory}, vol.~48, no.~3, pp. 695--707, 2002.

\bibitem{kirac1996results}
A.~Kirac and P.~Vaidyanathan, ``Results on lattice vector quantization with
  dithering,'' \emph{IEEE Transactions On Circuits and Systems II: Analog and
  Digital Signal Processing}, vol.~43, no.~12, pp. 811--826, 1996.

\bibitem{wilson2000layered}
D.~B. Wilson, ``Layered multishift coupling for use in perfect sampling
  algorithms (with a primer on {CFTP}),'' \emph{Monte Carlo Methods}, vol.~26,
  pp. 141--176, 2000.

\bibitem{hegazy2022randomized}
M.~Hegazy and C.~T. Li, ``Randomized quantization with exact error
  distribution,'' in \emph{2022 IEEE Information Theory Workshop (ITW)}.\hskip
  1em plus 0.5em minus 0.4em\relax IEEE, 2022, pp. 350--355.

\bibitem{Dwork06DP}
C.~Dwork, F.~McSherry, K.~Nissim, and A.~Smith, ``Calibrating noise to
  sensitivity in private data analysis,'' in \emph{Theory of
  Cryptography}.\hskip 1em plus 0.5em minus 0.4em\relax Springer Berlin
  Heidelberg, 2006, pp. 265--284.

\bibitem{Dwork14Book}
\BIBentryALTinterwordspacing
C.~Dwork and A.~Roth, ``The algorithmic foundations of differential privacy,''
  \emph{Found. Trends Theor. Comput. Sci.}, vol.~9, no. 3-4, pp. 211--407,
  2014. [Online]. Available: \url{https://doi.org/10.1561/0400000042}
\BIBentrySTDinterwordspacing

\bibitem{duchi2013local}
J.~C. Duchi, M.~I. Jordan, and M.~J. Wainwright, ``Local privacy and
  statistical minimax rates,'' in \emph{2013 IEEE 54th Ann. IEEE Symp.
  Found.}\hskip 1em plus 0.5em minus 0.4em\relax IEEE, 2013, pp. 429--438.

\bibitem{el2022differential}
A.~El~Ouadrhiri and A.~Abdelhadi, ``Differential privacy for deep and federated
  learning: A survey,'' \emph{IEEE access}, vol.~10, pp. 22\,359--22\,380,
  2022.

\bibitem{li2020federated}
T.~Li, A.~K. Sahu, A.~Talwalkar, and V.~Smith, ``Federated learning:
  Challenges, methods, and future directions,'' \emph{IEEE signal processing
  magazine}, vol.~37, no.~3, pp. 50--60, 2020.

\bibitem{lang2022joint}
N.~Lang and N.~Shlezinger, ``Joint privacy enhancement and quantization in
  federated learning,'' in \emph{2022 IEEE International Symposium on
  Information Theory (ISIT)}.\hskip 1em plus 0.5em minus 0.4em\relax IEEE,
  2022, pp. 2040--2045.

\bibitem{shahmiri2023}
A.~M. Shahmiri, C.~W. Ling, and C.~T. Li, ``Communication-efficient laplace
  mechanism for differential privacy via random quantization,'' \emph{arXiv
  preprint cs.CR/2309.06982}, 2023.

\bibitem{hasircioglu2023communication}
B.~Hasircioglu and D.~Gunduz, ``Communication efficient private federated
  learning using dithering,'' \emph{arXiv preprint arXiv:2309.07809}, 2023.

\bibitem{hegazy2023compression}
M.~Hegazy, R.~Leluc, C.~T. Li, and A.~Dieuleveut, ``Compression with exact
  error distribution for federated learning,'' \emph{arXiv preprint
  cs.LG/2310.20682}, 2023.

\bibitem{yan2023layered}
G.~Yan, T.~Li, T.~Lan, K.~Wu, and L.~Song, ``Layered randomized quantization
  for communication-efficient and privacy-preserving distributed learning,''
  \emph{arXiv preprint arXiv:2312.07060}, 2023.

\bibitem{agustsson2020universally}
E.~Agustsson and L.~Theis, ``Universally quantized neural compression,''
  \emph{Advances in neural information processing systems}, vol.~33, pp.
  12\,367--12\,376, 2020.

\bibitem{flamich2020compressing}
G.~Flamich, M.~Havasi, and J.~M. Hern{\'a}ndez-Lobato, ``Compressing images by
  encoding their latent representations with relative entropy coding,''
  \emph{Advances in Neural Information Processing Systems}, vol.~33, pp.
  16\,131--16\,141, 2020.

\bibitem{shlezinger2020uveqfed}
N.~Shlezinger, M.~Chen, Y.~C. Eldar, H.~V. Poor, and S.~Cui, ``{UVeQFed}:
  Universal vector quantization for federated learning,'' \emph{IEEE
  Transactions on Signal Processing}, vol.~69, pp. 500--514, 2020.

\bibitem{yang2022introduction}
Y.~Yang, S.~Mandt, and L.~Theis, ``An introduction to neural data
  compression,'' \emph{arXiv preprint arXiv:2202.06533}, 2022.

\bibitem{vectquant_isit}
C.~W. Ling and C.~T. Li, ``Vector quantization with error uniformly distributed
  over an arbitrary set,'' in \emph{Proc. IEEE Int. Symp. Inf. Theory}, June
  2023.

\bibitem{andres2013geo}
M.~E. Andr{\'e}s, N.~E. Bordenabe, K.~Chatzikokolakis, and C.~Palamidessi,
  ``Geo-indistinguishability: Differential privacy for location-based
  systems,'' in \emph{Proceedings of the 2013 ACM SIGSAC conference on Computer
  \& communications security}, 2013, pp. 901--914.

\bibitem{reimherr2019elliptical}
M.~Reimherr and J.~Awan, ``Elliptical perturbations for differential privacy,''
  \emph{Advances in Neural Information Processing Systems}, vol.~32, 2019.

\bibitem{gray1993dithered}
R.~M. Gray and T.~G. Stockham, ``Dithered quantizers,'' \emph{IEEE Transactions
  on Information Theory}, vol.~39, no.~3, pp. 805--812, 1993.

\bibitem{Lipshitz1992}
S.~Lipshitz, R.~Wannamaker, and J.~Vanderkooy, ``Quantization and dither: A
  theoretical survey,'' \emph{Journal of the Audio Engineering Society},
  vol.~40, pp. 355--374, 05 1992.

\bibitem{Ziv1985}
J.~Ziv, ``On universal quantization,'' \emph{IEEE Transactions on Information
  Theory}, vol.~31, no.~3, pp. 344--347, 1985.

\bibitem{Gray1990spectra}
R.~Gray, ``Quantization noise spectra,'' \emph{IEEE Transactions on Information
  Theory}, vol.~36, no.~6, pp. 1220--1244, 1990.

\bibitem{Zamir1992}
R.~Zamir and M.~Feder, ``On universal quantization by randomized
  uniform/lattice quantizers,'' \emph{IEEE Transactions on Information Theory},
  vol.~38, no.~2, pp. 428--436, 1992.

\bibitem{Zamir1995RD}
------, ``Rate-distortion performance in coding bandlimited sources by sampling
  and dithered quantization,'' \emph{IEEE Transactions on Information Theory},
  vol.~41, no.~1, pp. 141--154, 1995.

\bibitem{Zamir1996}
------, ``Information rates of pre/post-filtered dithered quantizers,''
  \emph{IEEE Transactions on Information Theory}, vol.~42, no.~5, pp.
  1340--1353, 1996.

\bibitem{Mashiach2013noiseShaping}
A.~Mashiach and R.~Zamir, ``Noise-shaped quantization for nonuniform
  sampling,'' in \emph{2013 IEEE International Symposium on Information
  Theory}, 2013, pp. 1187--1191.

\bibitem{Akyol2012}
E.~Akyol and K.~Rose, ``On constrained randomized quantization,'' \emph{IEEE
  Transactions on Signal Processing}, vol.~61, 06 2012.

\bibitem{bennett2002entanglement}
C.~H. Bennett, P.~W. Shor, J.~Smolin, and A.~V. Thapliyal,
  ``Entanglement-assisted capacity of a quantum channel and the reverse
  {S}hannon theorem,'' \emph{IEEE Trans. Inf. Theory}, vol.~48, no.~10, pp.
  2637--2655, 2002.

\bibitem{winter2002compression}
A.~Winter, ``Compression of sources of probability distributions and density
  operators,'' \emph{arXiv preprint quant-ph/0208131}, 2002.

\bibitem{cuff2013distributed}
P.~Cuff, ``Distributed channel synthesis,'' \emph{IEEE Trans. Inf. Theory},
  vol.~59, no.~11, pp. 7071--7096, Nov 2013.

\bibitem{Harsha2007}
P.~Harsha, R.~Jain, D.~McAllester, and J.~Radhakrishnan, ``The communication
  complexity of correlation,'' in \emph{Twenty-Second Annual IEEE Conference on
  Computational Complexity (CCC'07)}, 2007, pp. 10--23.

\bibitem{braverman2014public}
M.~Braverman and A.~Garg, ``Public vs private coin in bounded-round
  information,'' in \emph{International Colloquium on Automata, Languages, and
  Programming}.\hskip 1em plus 0.5em minus 0.4em\relax Springer, 2014, pp.
  502--513.

\bibitem{li2018universal}
C.~T. {Li} and A.~{El Gamal}, ``A universal coding scheme for remote generation
  of continuous random variables,'' \emph{IEEE Trans. Inf. Theory}, vol.~64,
  no.~4, pp. 2583--2592, April 2018.

\bibitem{sfrl_trans}
C.~T. Li and A.~{El Gamal}, ``Strong functional representation lemma and
  applications to coding theorems,'' \emph{IEEE Trans. Inf. Theory}, vol.~64,
  no.~11, pp. 6967--6978, Nov 2018.

\bibitem{kumar2014exact}
G.~R. Kumar, C.~T. Li, and A.~{El Gamal}, ``Exact common information,'' in
  \emph{Proc. IEEE Int. Symp. Inf. Theory}, June 2014, pp. 161--165.

\bibitem{li2017distributed}
C.~T. Li and A.~El~Gamal, ``Distributed simulation of continuous random
  variables,'' \emph{IEEE Trans. Inf. Theory}, vol.~63, no.~10, pp. 6329--6343,
  2017.

\bibitem{chou1989entropy}
P.~A. Chou, T.~Lookabaugh, and R.~M. Gray, ``Entropy-constrained vector
  quantization,'' \emph{IEEE Transactions on acoustics, speech, and signal
  processing}, vol.~37, no.~1, pp. 31--42, 1989.

\bibitem{frink1933jordan}
O.~Frink, ``Jordan measure and {R}iemann integration,'' \emph{Annals of
  Mathematics}, pp. 518--526, 1933.

\bibitem{robert2004monte}
C.~P. Robert and G.~Casella, ``Monte {C}arlo statistical methods,''
  \emph{Springer Texts in Statistics}, p. 274, 2004.

\bibitem{li2010distribution}
M.~Li, J.~Klejsa, and W.~B. Kleijn, ``Distribution preserving quantization with
  dithering and transformation,'' \emph{IEEE Signal Processing Letters},
  vol.~17, no.~12, pp. 1014--1017, 2010.

\bibitem{Wendel1948Note}
J.~G. Wendel, ``Note on the {G}amma function,'' \emph{American Mathematical
  Monthly}, vol.~55, p. 563, 1948.

\bibitem{qi2009bounds}
F.~Qi, ``Bounds for the ratio of two gamma functions--from wendel's and related
  inequalities to logarithmically completely monotonic functions,'' \emph{arXiv
  preprint math.CA/0904.1048}, 2009.

\end{thebibliography}

\ifshortver
\else

% \clearpage
\appendices

\section{Examples of Nyquist-$\mathbf{G}$ Distributions}\label{sec:nyquist}

\begin{figure}
    \centering
    \includegraphics[width=400pt]{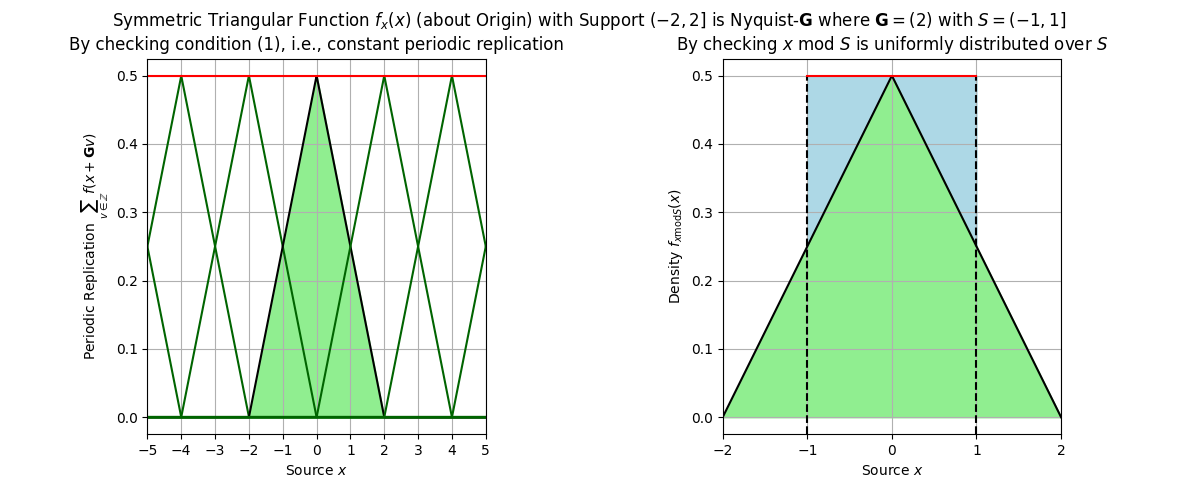}
    \caption{Example of a Nyquist-$\mathbf{G}$ distribution. 
    Consider a one-dimensional distribution with density $f_{x}(x)=\max\{1/2 - |x|/4,\, 0\}$ (green-shaded area), which is Nyquist-$\mathbf{G}$ with respect to the lattice $\Lambda(\mathbf{G})=2\mathbb{Z}$.
    The left figure shows that the periodic replication with respect to the lattice $\Lambda(\mathbf{G})$ is equal to a constant function (the red line).
    The right figure shows that taking modulo reduction with respect to the basic cell $S=(-1,1]$ of $\Lambda(\mathbf{G})$ on $f_{x}$, i.e., taking the distribution of $x\;\mathrm{mod}\;S$ where $x \sim f_{x}$, will give a uniform distribution over $S$ (the blue rectangle).
    % light blue-shaded area and the green-shaded area under the red constant curve with support $S$).
    % Notice that the projection of the red constant curve on the basic cell $S$ in the left figure is equal to the red constant curve on the right figure.
    }
    \label{fig::NyquistGDist}
\end{figure}

% \textcolor{blue}{
% the distribution of the modulo-$\Lambda(\mathbf{G})$ reduction of a random source $\mathbf{x}$, is a uniform distribution over a basic cell $S$ of $\Lambda(\mathbf{G})$ \cite[Equation (4.17)]{zamir2014}.
% The geometric intuition underlying the equivalence between condition (\ref{eq:NyquistG}) and the uniformity of the folded distribution of $\mathbf{x}$ is based on the observation that taking modulo-$\Lambda(\mathbf{G})$ reduction of $\mathbf{x}$ is equivalent to the projection of the periodic replication of the density $f$ onto a basic cell of $\Lambda(\mathbf{G})$.?????? 
% Please refer to Figure~\ref{fig::NyquistGDist} for an illustration.
% The proofs are omitted, but the interested readers can refer to \cite{kirac1996results,zamir2014}.}
%One example of Nyquist-$\mathbf{G}$ distributions is the uniform distribution over a basic cell.
We give several examples of Nyquist-$\mathbf{G}$  distributions below:
\begin{enumerate}
 \item One-dimensional triangular distribution with density $f_{x}(x)=\max\{1/\Delta - |x|/\Delta^2,\, 0\}$, $\Delta > 0$, which is Nyquist-$\mathbf{G}$ with respect to the lattice $\Lambda(\mathbf{G})=\Delta\mathbb{Z}$. See Figure~\ref{fig::NyquistGDist}.
 \item Uniform distribution over any basic cell of $\Lambda(\mathbf{G})$ \cite{kirac1996results,zamir2014}. 
 % Specifically, the uniform distribution over the Voronoi region $(-\frac{1}{2}, \frac{1}{2}]$ of the integer lattice $\mathbb{Z}$.
 \item Uniform distribution over a basic cell of a sublattice $\Lambda(\tilde{\mathbf{G}}) \subseteq \Lambda(\mathbf{G})$ \cite{zamir2014}.
 \item Piecewice uniform distribution over an arbitrary union of nonoverlapping basic cells of $\Lambda(\mathbf{G})$ \cite{kirac1996results}.
\end{enumerate} 
Interested readers are referred to \cite{kirac1996results,zamir2014} for more properties of Nyquist-$\mathbf{G}$ distributions.

\section{Proof of Proposition \ref{prop:quantize_nyquist}\label{subsec:quantize_nyquist_pf}}
Assume $\bar{f}_{Q}$ is defined using the basic cell $S=\{\mathbf{G}\mathbf{u}:\,\mathbf{u}\in [0,1)^n\}$. We will prove that the error distribution is always the same as this $\bar{f}_{Q}$ for any Nyquist-$\mathbf{G}$ distribution $f$. Letting $\mathbf{x} \sim f$, for any measurable $A\subseteq \mathbb{R}^n$,
% Letting $f$ be the probability density function of $\mathbf{x}$,
% we have
\begin{align*}
 & \mathbb{P}\left(\mathbf{x}-Q(\mathbf{x})\in A\right)\\
 & =\int\mathbf{1}\{\mathbf{x}-Q(\mathbf{x})\in A\}f(\mathbf{x})\mathrm{d}\mathbf{x}\\
 & =|\mathrm{det}\mathbf{G}|\sum_{\mathbf{v}\in\mathbb{Z}^{n}}\int_{[0,1)^{n}}\mathbf{1}\{\mathbf{G}(\mathbf{v}+\mathbf{u})-Q(\mathbf{G}(\mathbf{v}+\mathbf{u}))\in A\}\\
 & \qquad\qquad\qquad\;\;\;\;\;\cdot f(\mathbf{G}(\mathbf{v}+\mathbf{u}))\mathrm{d}\mathbf{u}\\
 & \stackrel{(a)}{=}|\mathrm{det}\mathbf{G}|\sum_{\mathbf{v}\in\mathbb{Z}^{n}}\int_{[0,1)^{n}}\!\!\!\!\mathbf{1}\{\mathbf{G}\mathbf{u}-Q(\mathbf{G}\mathbf{u})\in A\}f(\mathbf{G}(\mathbf{v}+\mathbf{u}))\mathrm{d}\mathbf{u}\\
 & =|\mathrm{det}\mathbf{G}|\!\int_{[0,1)^{n}}\!\!\!\!\!\!\!\!\mathbf{1}\{\mathbf{G}\mathbf{u}-Q(\mathbf{G}\mathbf{u})\in A\}\Big(\sum_{\mathbf{v}\in\mathbb{Z}^{n}}\! f(\mathbf{G}(\mathbf{v}+\mathbf{u}))\!\Big)\mathrm{d}\mathbf{u}\\
 & \stackrel{(b)}{=}\int_{[0,1)^{n}}\mathbf{1}\{\mathbf{G}\mathbf{u}-Q(\mathbf{G}\mathbf{u})\in A\}\mathrm{d}\mathbf{u}\\
 & =\int_{A}\bar{f}_{Q}(\mathbf{x})\mathrm{d}\mathbf{x},
\end{align*}
where (a) is by the shift-periodicity of $Q$ and (b) is by (\ref{eq:NyquistG}).
Hence $\mathbf{x}-Q(\mathbf{x}) \sim \bar{f}_{Q}$.

\section{Proof of Corollary \ref{cor:unif_approx}\label{subsec:unif_approx_pf}}

Without loss of generality, assume $\mathbf{G}=\mathbf{I}$ is the identity matrix.
For $A,B\subseteq\mathbb{R}^{n}$, write $A\ominus B:=\{\mathbf{x}\in\mathbb{R}^{n}:\,B+\mathbf{x}\subseteq A\}$
for the erosion of $A$ by $B$. By Lemma 1 in~\cite{li2017distributed} (note that any convex set is orthogonally convex by definition),
\[
\mu\left(A\ominus[0,1)^{n}\right)\ge\mu(A)-nr^{n-1}\nu_{n-1}.
\]
We have
\begin{align*}
 & \int_{[0,1)^{n}}\sum_{\mathbf{v}\in\mathbb{Z}^{n}}\mathbf{1}\{[0,1)^{n}+\mathbf{v}+\mathbf{u}\subseteq A\}\mathrm{d}\mathbf{u}\\
 & =\int_{\mathbb{R}^{n}}\mathbf{1}\{[0,1)^{n}+\mathbf{u}\subseteq A\}\mathrm{d}\mathbf{u}\\
 & =\mu\left(A\ominus[0,1)^{n}\right)\\
 & \ge\mu(A)-nr^{n-1}\nu_{n-1}.
\end{align*}
Hence there exists $\mathbf{u}_{0} \in [0,1)^n$ such that
\begin{align*}
 & \sum_{\mathbf{v}\in\mathbb{Z}^{n}}\mathbf{1}\{[0,1)^{n}+\mathbf{v}+\mathbf{u}_{0}\subseteq A\}  \ge\mu(A)-nr^{n-1}\nu_{n-1}.
\end{align*}
Let $C := \bigcup_{\mathbf{v}\in\mathbb{Z}^{n}:\,[0,1)^{n}+\mathbf{v}+\mathbf{u}_{0}\subseteq A}([0,1)^{n}+\mathbf{v}+\mathbf{u}_{0}) \subseteq A$. Note that $\mathrm{Unif}(C)$
is a Nyquist-$\mathbf{I}$ distribution, and its total variation distance
from $\mathrm{Unif}(A)$ is upper-bounded by
\begin{align*}
 &\delta_{\mathrm{TV}}(\mathrm{Unif}(C),\mathrm{Unif}(A)) \\
 &= \sup_{B \subseteq \mathbb{R}^n \; \mathrm{measurable}} \Big(\mathbb{P}_{\mathbf{x} \sim \mathrm{Unif}(A)}(\mathbf{x} \in B)-\mathbb{P}_{\mathbf{x} \sim \mathrm{Unif}(C)}(\mathbf{x} \in B)\Big) \\
 &=\mathbb{P}_{\mathbf{x} \sim \mathrm{Unif}(A)}(\mathbf{x} \in A\setminus C)-\mathbb{P}_{\mathbf{x} \sim \mathrm{Unif}(C)}(\mathbf{x} \in A\setminus C) \\
 &= 1-\frac{\mu(C)}{\mu(A)}\\
 & =1-\frac{\sum_{\mathbf{v}\in\mathbb{Z}^{n}}\mathbf{1}\{[0,1)^{n}+\mathbf{v}+\mathbf{u}_{0}\subseteq A\}}{\mu(A)}\\
 & \le\frac{nr^{n-1}\nu_{n-1}}{\mu(A)}.
\end{align*}
The result follows from Proposition~\ref{prop:quantize_nyquist}.

\section{Proof of Proposition \ref{prop:ent_large}\label{subsec:ent_large_pf}}
We first give the intuition of the proof. Our goal is to approximate $H(Q(\mathbf{x}))$ where $\mathbf{x} \sim \mathrm{Unif}(rA)$ is uniform over a large set $rA$. From Definition \ref{def:normalized_ent}, $H(Q(\mathbf{x})) = \bar{H}(Q)+\log |\det \mathbf{G}|$ when $\mathbf{x} \sim \mathrm{Unif}(T)$ is uniform over a preimage basic cell $T$. Therefore, when $\mathbf{x} \sim \mathrm{Unif}(\mathbf{G}C + T)$, where $C \subseteq \mathbb{Z}^n$ is finite (i.e., $\mathbf{x}$ is uniform over the union of lattice-translated copies of $T$), we have $H(Q(\mathbf{x})) = \bar{H}(Q) + \log |C| +\log |\det \mathbf{G}|$. The idea is to approximate $rA$ by unions of lattice-translated copies of $T$, by giving inner and outer bounds of $rA$ using such unions, which allows us to bound the entropy $H(Q(\mathbf{x}))$ where $\mathbf{x} \sim \mathrm{Unif}(rA)$.

We then present the proof. Assume $\mathbf{G}=\mathbf{I}$ is the identity matrix without loss of generality. Consider
the preimage basic cell $T=Q^{-1}([0,1)^{n})$. By the definition
of shift-periodic quantizer, we know $T$ is bounded, and hence we
can let $\eta:=2\sup_{\mathbf{x}\in T}\Vert\mathbf{x}\Vert$. Consider
$\mathbf{x}\sim\mathrm{Unif}(rA)$ where $A\subseteq\mathbb{R}^{n}$
has measure-zero boundary. 
Consider the erosion $A\ominus\gamma B_{n}=\{\mathbf{x}\in\mathbb{R}^{n}:\,\gamma B_{n}+\mathbf{x}\subseteq A\}$ for $0<\gamma <1$.
Since $A$ has measure-zero boundary, $\lim_{\gamma\to0}\mu(A\ominus\gamma B_{n})=\mu(A)$.
Consider 
\begin{align}
\underline{A}_{\gamma} &:=\bigcup_{\mathbf{v}\in\mathbb{Z}^{n}:\,\gamma(T+\mathbf{v})\subseteq A}\gamma(T+\mathbf{v}).\label{eq:A_under_gamma}
\end{align}
Note that if $\gamma\eta B_{n}+\mathbf{x}\subseteq A$, then any translated
copy of $\gamma T$ that contains $\mathbf{x}$ must also be included
in $A$, and hence there exists $\mathbf{v}\in\mathbb{Z}^{n}$ such
that $\mathbf{x}\in\gamma(T+\mathbf{v})\subseteq A$. Hence $A\ominus\gamma\eta B_{n}\subseteq\underline{A}_{\gamma}\subseteq A$,
and $\lim_{\gamma\to0}\mu(\underline{A}_{\gamma})=\mu(A)$, i.e., $\underline{A}_{\gamma}$ are inner bounds of $A$ that approach $A$. 
From \eqref{eq:A_under_gamma}, $\gamma^{-1}\underline{A}_{\gamma} = \{\mathbf{v}\in\mathbb{Z}^{n}:\,\gamma(T+\mathbf{v})\subseteq A\} + T$. Since there is no overlapping between $T+\mathbf{v}$ and $T+\mathbf{v}'$ for $\mathbf{v},\mathbf{v}' \in \mathbb{Z}^n$, $\mathbf{v} \neq \mathbf{v}'$, we know that 
if
$\mathbf{v}\sim\mathrm{Unif}\{\mathbf{v}\in\mathbb{Z}^{n}:\,\gamma(T+\mathbf{v})\subseteq A\}$ is
independent of $\mathbf{u}\sim\mathrm{Unif}(T)$, then
$\mathbf{v}+\mathbf{u}\sim\mathrm{Unif}(\gamma^{-1}\underline{A}_{\gamma})$, and
\begin{align}
 & H(Q(\mathbf{v}+\mathbf{u}))\nonumber \\
 & \stackrel{(a)}{=}H(\mathbf{v},Q(\mathbf{v}+\mathbf{u}))\nonumber \\
 & =H(Q(\mathbf{v}+\mathbf{u})|\mathbf{v})+\log\left|\{\mathbf{v}\in\mathbb{Z}^{n}:\,\gamma(T+\mathbf{v})\subseteq A\}\right|\nonumber \\
 & \stackrel{(b)}{=}H(Q(\mathbf{u}))+\log\mu(\gamma^{-1}\underline{A}_{\gamma})\nonumber \\
 & =\bar{H}(Q)+\log\mu(\underline{A}_{\gamma})-n\log\gamma,\label{eq:HQvu}
\end{align}
where (a) is because $Q(\mathbf{v}+\mathbf{u})\,\mathrm{mod}\,[0,1)^n=\mathbf{v}$
since $\mathbf{u}\in T=Q^{-1}([0,1)^n)$ and (b) is because the measure of $\gamma^{-1}\underline{A}_{\gamma}$ is $\mu(T)$ (which equals $\det \mathbf{G} = 1$ since $T$ is a basic cell) times the number of lattice points $\mathbf{v}\in\mathbb{Z}^{n}$ satisfying $\gamma(T+\mathbf{v})\subseteq A$. Let $\mathbf{x}\sim\mathrm{Unif}(\gamma^{-1}A)$.
We have
\begin{align*}
 &H(Q(\mathbf{x}))\\
 &\ge H(Q(\mathbf{x}),\,\mathbf{1}\{\mathbf{x}\in\gamma^{-1}\underline{A}_{\gamma}\})-H(\mathbf{1}\{\mathbf{x}\in\gamma^{-1}\underline{A}_{\gamma}\})\\
 &\ge \mathbb{P}(\mathbf{x}\in\gamma^{-1}\underline{A}_{\gamma}) H(Q(\mathbf{x})\,|\,\mathbf{x}\in\gamma^{-1}\underline{A}_{\gamma})-H(\mathbf{1}\{\mathbf{x}\in\gamma^{-1}\underline{A}_{\gamma}\}) \\
 & =\frac{\mu(\gamma^{-1}\underline{A}_{\gamma})}{\mu(\gamma^{-1}A)}H(Q(\mathbf{x})\,|\,\mathbf{x}\in\gamma^{-1}\underline{A}_{\gamma})-H_{b}\left(\frac{\mu(\gamma^{-1}\underline{A}_{\gamma})}{\mu(\gamma^{-1}A)}\right)\\
 & \stackrel{(c)}{=}\frac{\mu(\underline{A}_{\gamma})}{\mu(A)}\left(\bar{H}(Q)+\log\mu(\underline{A}_{\gamma})-n\log\gamma\right)-H_{b}\left(\frac{\mu(\underline{A}_{\gamma})}{\mu(A)}\right),
\end{align*}
where $H_b$ is the binary entropy function, and (c) is by (\ref{eq:HQvu}).
Since $\lim_{\gamma\to0}\mu(\underline{A}_{\gamma})=\mu(A)$, we have
\begin{align}
 & \underset{\gamma\to0}{\mathrm{liminf}}\left(H(Q(\mathbf{x}))-\log\mu(\gamma^{-1}A)\right)\nonumber \\
 & =\underset{\gamma\to0}{\mathrm{liminf}}\left(H(Q(\mathbf{x}))-\log\mu(A)+n\log\gamma\right)\nonumber \\
 & \ge\bar{H}(Q).\label{eq:liminf}
\end{align}
For the other direction, consider the dilation $A\oplus\gamma B_{n}=\{\mathbf{x}\in\mathbb{R}^{n}:\,\gamma B_{n}+\mathbf{x}\cap A\neq\emptyset\}$.
Since $A$ has measure-zero boundary, $\lim_{\gamma\to0}\mu(A\oplus\gamma B_{n})=\mu(A)$.
Consider 
\[
\overline{A}_{\gamma}:=\bigcup_{\mathbf{v}\in\mathbb{Z}^{n}:\,\gamma(T+\mathbf{v})\cap A\neq\emptyset}\gamma(T+\mathbf{v}).
\]
Note that if $\mathbf{x}\in\gamma(T+\mathbf{v})\cap A$, then $\gamma(T+\mathbf{v})\subseteq\gamma\eta B_{n}+\mathbf{x}$,
and hence $A\subseteq\overline{A}_{\gamma}\subseteq A\oplus\gamma\eta B_{n}$,
and $\lim_{\gamma\to0}\mu(\overline{A}_{\gamma})=\mu(A)$\, i.e., $\overline{A}_{\gamma}$ are outer bounds of $A$ that approach $A$. 
If
$\mathbf{v}\sim\mathrm{Unif}\{\mathbf{v}\in\mathbb{Z}^{n}:\,\gamma(T+\mathbf{v})\cap A\neq\emptyset\}$ is
independent of $\mathbf{u}\sim\mathrm{Unif}(T)$, then by the same arguments as  \eqref{eq:HQvu}, we have $\mathbf{v}+\mathbf{u}\sim\mathrm{Unif}(\gamma^{-1}\overline{A}_{\gamma})$, and
% \textcolor{blue}{By the similar argument as above for independent $\mathbf{v}$ and $\mathbf{u}$, w}e have $\mathbf{v}+\mathbf{u}\sim\mathrm{Unif}(\gamma^{-1}\overline{A}_{\gamma})$,
% and 
\[
H(Q(\mathbf{v}+\mathbf{u}))=\bar{H}(Q)+\log\mu(\overline{A}_{\gamma})-n\log\gamma.
\]
% by the same steps as \eqref{eq:HQvu}. 
Let $\mathbf{x}\sim\mathrm{Unif}(\gamma^{-1}A)$.
We have
\begin{align*}
 & H(Q(\mathbf{v}+\mathbf{u}))\\
 & \ge H(Q(\mathbf{v}\!+\!\mathbf{u}),\,\mathbf{1}\{\mathbf{v}\!+\!\mathbf{u}\in\gamma^{-1}A\})-H(\mathbf{1}\{\mathbf{v}\!+\!\mathbf{u}\in\gamma^{-1}A\})\\
 & \ge\frac{\mu(A)}{\mu(\overline{A}_{\gamma})}H(Q(\mathbf{v}+\mathbf{u})\,|\,\mathbf{v}+\mathbf{u}\in\gamma^{-1}A)-H_{b}\left(\frac{\mu(A)}{\mu(\overline{A}_{\gamma})}\right)\\
 & =\frac{\mu(A)}{\mu(\overline{A}_{\gamma})}H(Q(\mathbf{x}))-H_{b}\left(\frac{\mu(A)}{\mu(\overline{A}_{\gamma})}\right).
\end{align*}
Hence,
\begin{align*}
 & H(Q(\mathbf{x}))\\
 & \le\frac{\mu(\overline{A}_{\gamma})}{\mu(A)}\bigg(\bar{H}(Q)+\log\mu(\overline{A}_{\gamma})-n\log\gamma+H_{b}\bigg(\frac{\mu(A)}{\mu(\overline{A}_{\gamma})}\bigg)\bigg).
\end{align*}
We have
\begin{align*}
 & \underset{\gamma\to0}{\mathrm{limsup}}\left(H(Q(\mathbf{x}))-\log\mu(\gamma^{-1}A)\right)\le\bar{H}(Q).
\end{align*}
Combining with \eqref{eq:liminf}, we have $\lim_{\gamma\to0}(H(Q(\mathbf{x}))-\log\mu(\gamma^{-1}A))=\bar{H}(Q)$.

\section{Proof of Proposition \ref{prop:lowerbound}\label{subsec:lowerbound_pf}}
Let $\mathbf{G}$ be the generator matrix and $\mathbf{x}\sim \mathrm{Unif}(T)$ where $T$ is the preimage basic cell of $Q$.
%Consider the Markov chain 
%\[\mathbf{x} \; \mathrm{mod} \; T \longleftrightarrow \mathbf{x} %\longleftrightarrow \mathrm{frac}\big(\mathbf{G}^{-1}Q(\mathbf{x})\big).\]
%\textcolor{cyan}{$\mathbf{x} \; \mathrm{mod}$ and $\mathrm{frac}\big(\mathbf{G}^{-1}Q(\mathbf{x})\big)$ are functions of $\mathbf{x}$, so this is trivially a Markov chain? Why do we need Markov chain here?}
Notice that
\begin{align*}
\bar{H}(Q)&=H(Q(\mathbf{x}))- \log |\det \mathbf{G}| \\
&\stackrel{(a)}=I(\mathbf{x};Q(\mathbf{x}))- \log |\det \mathbf{G}| \\
&=h(\mathbf{x})-h(\mathbf{x}|Q(\mathbf{x}))- \log |\det \mathbf{G}| \\
&\stackrel{(b)}{=}-h(\mathbf{x}|Q(\mathbf{x})) \\
&=-h(\mathbf{x}-Q(\mathbf{x})|Q(\mathbf{x})) \\
&\geq -h(\mathbf{x}-Q(\mathbf{x})) \\
&\stackrel{(c)}{=}-h(\bar{f}_{Q}),
\end{align*}
where (a) is because $H(Q(\mathbf{x})|\mathbf{x})=0$, (b) is because $h(\mathbf{x})=\log|\det \mathbf{G}|$, and (c) is by Proposition~\ref{prop:quantize_nyquist}.

\section{Proof of Lemma \ref{lem:dissect}\label{subsec:dissect_pf}}

We first prove the following claim: For $\tilde{A},\tilde{S}\subseteq\mathbb{R}^{n}$
sets with finite volume and measure-zero boundaries, and 
\begin{equation} \label{inq:phi_z_ub}
0<t\le \mu(\tilde{A})\mu(\tilde{S})/\mu(\tilde{S}-\tilde{A}),
\end{equation}
there exists $\mathbf{z}\in\mathbb{R}^{n}$ such that $\mu((\tilde{A}+\mathbf{z})\cap\tilde{S})=t$.
To prove this, note that 
\begin{align}
    \phi(\mathbf{z}):=\mu((\tilde{A}+\mathbf{z})\cap\tilde{S}) \label{eq:phi_z}
\end{align}
is a continuous function due to the measure-zero boundaries,\footnote{Note that $\phi(\mathbf{z}_1)-\phi(\mathbf{z}_2) \le \mu(((\tilde{A}+\mathbf{z}_1)\cap\tilde{S})\backslash ((\tilde{A}+\mathbf{z}_2)\cap\tilde{S})) \le \mu((\tilde{A}+\mathbf{z}_1-\mathbf{z}_2)\backslash \tilde{A}) \le \mu((\tilde{A}+\Vert \mathbf{z}_1-\mathbf{z}_2 \Vert B_n)\backslash \tilde{A}) = \mu(\tilde{A}+\Vert \mathbf{z}_1-\mathbf{z}_2 \Vert B_n) - \mu(\tilde{A})$. As $\Vert \mathbf{z}_1-\mathbf{z}_2 \Vert \to 0$, $\mu(\tilde{A}+\Vert \mathbf{z}_1-\mathbf{z}_2 \Vert B_n) \to \mu(\mathrm{cl}(\tilde{A}))=\mu(\tilde{A})$ since $\tilde{A}$ has measure-zero boundary.} 
and
\begin{align*}
 \int_{\tilde{S}-\tilde{A}}\phi(\mathbf{z})\mathrm{d}\mathbf{z}
 & =\int_{\mathbb{R}^{n}}\phi(\mathbf{z})\mathrm{d}\mathbf{z}\\
 & =\int_{\mathbb{R}^{n}}\int_{\tilde{S}}1\{\mathbf{x}\in\tilde{A}+\mathbf{z}\}\mathrm{d}\mathbf{x}\mathrm{d}\mathbf{z}\\
 & =\int_{\mathbb{R}^{n}}\int_{\tilde{S}}1\{\mathbf{x}-\mathbf{z}\in\tilde{A}\}\mathrm{d}\mathbf{x}\mathrm{d}\mathbf{z}\\
&=\int_{\tilde{S}}\int_{\mathbb{R}^{n}}1\{\mathbf{x}-\mathbf{z}\in\tilde{A}\}\mathrm{d}\mathbf{z}\mathrm{d}\mathbf{x} \\
 & =\mu(\tilde{S})\int_{\mathbb{R}^{n}}1\{\mathbf{x}\in\tilde{A}\}\mathrm{d}\mathbf{x} \; =\mu(\tilde{S})\mu(\tilde{A}).
\end{align*}
Hence, there exists $\mathbf{z}$ attaining the average value of $\phi(\mathbf{z})$
over $\tilde{S}-\tilde{A}$, which is $\mu(\tilde{A})\mu(\tilde{S})/\mu(\tilde{S}-\tilde{A})$.\footnote{In practice, we find $\mathbf{z}$ that maximizes $\phi(\mathbf{z})$ in \eqref{eq:phi_z} instead of attaining the average, in order to maximize the sizes of the larger components of the partition, and minimize the entropy. We consider the average in the proof only to simplify the analysis.}
Since $\phi(\mathbf{z})$ can be arbitrarily small, by the intermediate value theorem and the continuity of $\phi$,
we can find $\mathbf{z}$ with $\phi(\mathbf{z})=t$ for any $0<t\le\mu(\tilde{A})\mu(\tilde{S})/\mu(\tilde{S}-\tilde{A})$.

In the remainder of the proof, we apply this claim recursively. Intuitively, we first apply the claim on $A,S$ to obtain $\mathbf{z}_1$ and $T_1 = (A+\mathbf{z}_1)\cap S$ where $\mu(T_1)$ is  
close to the upper bound in \eqref{inq:phi_z_ub}.
% \textcolor{blue}{large with respect to Inequality~\eqref{inq:phi_z_ub}, i.e., obtaining $\mathbf{z}_{1}$ so that $\mu(T_{1})$ has value as close as the upper bound of Inequality \eqref{inq:phi_z_ub} as possible}. 
We then take $F(\mathbf{x})=\mathbf{z}_1$ if $\mathbf{x} \in T_1$. If $\mathbf{x} \sim \mathrm{Unif}(T_1)$, then $\mathbf{x}-F(\mathbf{x}) \in A$. The reason for requiring a large enough $\mu(T_1)$ is to maximize the size of the largest quantization cell, reducing the normalized entropy. We then apply the claim on $A \backslash (T_1-\mathbf{z}_1)$ and $S \backslash T_1$ to obtain $\mathbf{z}_2$, etc. 
% Refer to 
% \ifshortver
% \cite{vector_arxiv}
% \else
% Appendix~\ref{subsec:dissect_pf}
% \fi
% for the remainder of the proof.

We now present the precise arguments. Apply the claim recursively to obtain $\mathbf{z}_{1},\mathbf{z}_{2},\ldots\in\mathbb{R}^{n}$,
$T_{1},T_{2},\ldots\subseteq\mathbb{R}^{n}$, where $\mathbf{z}_{i}$
is obtained by applying the claim on $\tilde{A}_{i}:=A\backslash\bigcup_{j=1}^{i-1}(T_{j}-\mathbf{z}_{j})$
and $\tilde{S}_{i}:=S\backslash\bigcup_{j=1}^{i-1}T_{j}$, with the
requirement
\[
\mu\left((\tilde{A}_{i}+\mathbf{z}_{i})\cap\tilde{S}_{i}\right)=\frac{\mu(\tilde{A}_{i})\mu(\tilde{S}_{i})}{\mu(S-A)}\le\frac{\mu(\tilde{A}_{i})\mu(\tilde{S}_{i})}{\mu(\tilde{S}_{i}-\tilde{A}_{i})},
\]
and $T_{i}:=(\tilde{A}_{i}+\mathbf{z}_{i})\cap\tilde{S}_{i}$. Note
that $T_{1},T_{2},\ldots$ are disjoint, and $T_{1}-\mathbf{z}_{1},T_{2}-\mathbf{z}_{2},\ldots$
are disjoint. Hence $\mu(\tilde{A}_{i})=\mu(\tilde{S}_{i})=\mu(A)-\sum_{j=1}^{i-1}\mu(T_{j})$.
Letting $\gamma_{i}:=\mu(A)-\sum_{j=1}^{i}\mu(T_{j})$, we have
\begin{align*}
\gamma_{i}-\gamma_{i+1} & =\mu(T_{i+1}) =\frac{\mu(\tilde{A}_{i+1})\mu(\tilde{S}_{i+1})}{\mu(S-A)} =\frac{\gamma_{i}^{2}}{\mu(S-A)},
\end{align*}
and hence $\gamma_{0},\gamma_{1},\ldots$ follow the recurrence $\gamma_{0}=\mu(A)$,
$\gamma_{i+1}=\gamma_{i}-\gamma_{i}^{2}/\mu(S-A)$.

Let $\eta:=\mu(S-A)$, $\kappa:=\eta/\mu(A)$. We will prove inductively
that 
\begin{align}
\gamma_{i} & \le\frac{\eta}{\kappa+i}. \label{eq:gamma_bound}
\end{align}
The claim is clearly true for $i=0$. For $i=1$, 
\begin{align*}
\gamma_{i} & =\gamma_{0}-\frac{\gamma_{0}^{2}}{\eta} =\mu(A)\left(1-\frac{\mu(A)}{\eta}\right) \\
& \;\;\;\le\mu(A)\frac{1}{1+\mu(A)/\eta} =\frac{\eta}{\eta/\mu(A)+1} =\frac{\eta}{\kappa+i}.
% \gamma_{i} & =\gamma_{0}-\frac{\gamma_{0}^{2}}{\eta}\\
%  & =\mu(A)\left(1-\frac{\mu(A)}{\eta}\right)\\
%  & \le\mu(A)\frac{1}{1+\mu(A)/\eta}\\
%  & \le\frac{\eta}{\eta/\mu(A)+1}\\
%  & =\frac{\eta}{\kappa+i}.
\end{align*}
We then prove the claim for $i+1$ assuming that the claim is true
for $i$, where $i\ge1$. If $\gamma_{i}\le\eta/2$, since $\gamma-\gamma^{2}/\eta$
is nondecreasing in $\gamma$ for $\gamma\le\eta/2$,
\begin{align*}
\gamma_{i+1} & =\gamma_{i}-\frac{\gamma_{i}^{2}}{\eta}  \le\frac{\eta}{\kappa+i}\left(1-\frac{1}{\kappa+i}\right)  \le\frac{\eta}{\kappa+i+1}.
\end{align*}
If $\gamma_{i}>\eta/2$,
\begin{align*}
\gamma_{i+1} & =\gamma_{i}-\frac{\gamma_{i}^{2}}{\eta}\\
 & \le\frac{\eta}{\kappa+i}-\frac{(\eta/2)^{2}}{\eta}\\
 & =\eta\left(\frac{1}{\kappa+i}-\frac{1}{4}\right)\\
 & =\eta\left(\frac{1}{\kappa+i+1}+\frac{1}{(\kappa+i)(\kappa+i+1)}-\frac{1}{4}\right)\\
 & \le\frac{\eta}{\kappa+i+1}.
\end{align*}
Hence, $\gamma_{i}\le\eta/(\kappa+i)$ for $i\ge0$.

Since $\gamma_{i}\to0$ as $i\to\infty$, $T_{1},T_{2},\ldots$ form
a partition of $S$ (up to a null set), and $T_{1}-\mathbf{z}_{1},T_{2}-\mathbf{z}_{2},\ldots$
form a partition of $A$ (up to a null set). The function $F:S\to\mathbb{R}^{n}$
is defined as $F(\mathbf{x}):=\mathbf{z}_{i}$ where $i$ satisfies
$\mathbf{x}\in T_{i}$. Assign arbitrary $F(\mathbf{x})$ if no such
$i$ exists (which happens only in a null set). Consider the distribution
of $\mathbf{u}-F(\mathbf{u})$ where $\mathbf{u}\sim\mathrm{Unif}(S)$.
Conditional on the event $\mathbf{u}\in T_{i}$, we have $F(\mathbf{u})=\mathbf{z}_i$ by definition, and the conditional distribution of $\mathbf{u}$ is $\mathrm{Unif}(T_{i})$, and thus $\mathbf{u}-F(\mathbf{u})=\mathbf{u}-\mathbf{z}_{i}$
is uniform over $T_{i}-\mathbf{z}_{i}$.
% \textcolor{blue}{(the distribution $\mathrm{Unif}(T_i)$ shifted by $\mathbf{z}_i$)}.
Hence, when $\mathbf{u}\sim\mathrm{Unif}(S)$,
we have $\mathbf{u}-F(\mathbf{u})\sim\mathrm{Unif}(A)$.

Write $\ell(a):= -a \log a$. To compute the entropy, since $\sum_{j=1}^{i}\mu(T_{i})=\mu(A)-\gamma_{i}\ge\mu(A)-\eta/(\kappa+i)$,
by the Schur concavity of entropy,
\begin{align*}
 H(F(\mathbf{u}))
 & \le \sum_{i=1}^{\infty}\ell\bigg(\frac{\mu(T_{i})}{\mu(S)}\bigg)\\
 & =\sum_{i=1}^{\infty}\ell\bigg(\frac{\gamma_{i-1}-\gamma_{i}}{\mu(S)}\bigg)\\
 & \le\sum_{i=1}^{\infty}\ell\bigg(\frac{\eta/(\kappa+i-1)-\eta/(\kappa+i)}{\mu(S)}\bigg)\\
 & \stackrel{(a)}{\le}\frac{2\log e}{e}+\sum_{i=3}^{\infty}\ell\bigg(\frac{\eta/(\kappa+i-1)-\eta/(\kappa+i)}{\mu(S)}\bigg)\\
 & =\frac{2\log e}{e}+\sum_{i=3}^{\infty}\ell\left(\frac{\kappa}{\kappa+i-1}-\frac{\kappa}{\kappa+i}\right)\\
 & \stackrel{(b)}{\le}\frac{2\log e}{e}+\sum_{i=3}^{\infty}\ell\bigg(\frac{\kappa}{(\kappa+i-1)^{2}}\bigg)\\
 & \stackrel{(c)}{\le}\frac{2\log e}{e}+\sum_{i=3}^{\infty}\int_{\kappa+i-2}^{\kappa+i-1}\ell\Big(\frac{\kappa}{t^{2}}\Big)\mathrm{d}t\\
 & =\frac{2\log e}{e}+\int_{\kappa+1}^{\infty}\ell\Big(\frac{\kappa}{t^{2}}\Big)\mathrm{d}t\\
 & \le\frac{2\log e}{e}+\int_{\kappa}^{\infty}\ell\Big(\frac{\kappa}{t^{2}}\Big)\mathrm{d}t\\
 & =\frac{2\log e}{e}+\log(e^{2}\kappa)\\
 & \le\log\frac{\mu(S-A)}{\mu(A)}+4\;\mathrm{bits},
\end{align*}
where (a) is because $\ell(a)$ is maximized at $a=1/e$;
(b) is because 
\begin{align*}
\frac{\kappa}{\kappa+i-1}-\frac{\kappa}{\kappa+i} & =\frac{\kappa}{(\kappa+i-1)(\kappa+i)} \le\frac{\kappa}{(\kappa+2)^{2}}  \le\frac{1}{e}
\end{align*}
for $i\ge3$ since $\kappa\ge1$, and $\ell(a)$ is increasing
for $0\le a\le1/e$; and (c) is because $\kappa/t^{2}\le1/4\le1/e$
for $t\ge\kappa+i-2$, and $\ell(a)$ is increasing for $0\le a\le1/e$. 

For the second claim, note that if we sample $\mathbf{u} \sim \mathrm{Unif}(S)$, then by \eqref{eq:gamma_bound}, at $k$-th iteration, 
\begin{align*}
\mathbb{P}\bigg(\mathbf{u} \notin \bigcup_{j=1}^{k} T_j\bigg) 
&= \mu\bigg(S \big\backslash \bigcup_{j=1}^{k} T_j\bigg) / \mu(A)\\
&= \frac{\gamma_k}{\mu(A)} \\
&\le \frac{\eta}{\mu(A)(\kappa + k)} \\
&= \frac{\mu(S-A)}{\mu(S-A) + k \mu(A)}. \\
\end{align*}
Therefore, if we terminate this construction at the $k$-th iteration, i.e., we take $\tilde{F}(\mathbf{x}):=\mathbf{z}_i$ where $i \in \{1,\ldots,k\}$ satisfies $\mathbf{x} \in T_i$, or $\tilde{F}(\mathbf{x}):=\mathbf{z}_k$ if no such $i$ exists, then $\tilde{F}$ is the same as $F$ with large probability, i.e., $\mathbb{P}(\tilde{F}(\mathbf{u}) \neq F(\mathbf{u})) \le \mu(S-A)/(\mu(S-A) + k \mu(A))$. This implies that $\mathbf{u}-\tilde{F}(\mathbf{u})$ is approximately uniform over $A$, with the following bound on the total variation distance
\begin{align*}
\delta_{\mathrm{TV}}\big(\mathbf{u}-\tilde{F}(\mathbf{u}),\, \mathrm{Unif}(A)\big) \le \frac{\mu(S-A)}{\mu(S-A) + k \mu(A)}.
\end{align*}
We have the same bound on the entropy 
$H(\tilde{F}(\mathbf{u}))\le\log\frac{\mu(S-A)}{\mu(A)}+4$ since terminating the construction early can only reduce the entropy.
This is useful if we require the number of quantization cells to be finite and bounded by $k$.

\section{Proof of Corollary \ref{cor:ball}\label{subsec:cor_ball_pf}}

Consider the lattice generated by $\mathbf{G}=(\mu(B_{n}))^{1/n}\mathbf{I}$
with a basic cell $S=(\mu(B_{n}))^{1/n}[-1/2,1/2)^{n}$. We have $S\subseteq(\sqrt{n}/2)(\mu(B_{n}))^{1/n}B_{n}$,
and hence 
\[
S-B_{n}\subseteq\left(\frac{\sqrt{n}}{2}(\mu(B_{n}))^{1/n}+1\right)B_{n}.
\]
By Theorem~\ref{thm:quantization},
\begin{align*}
\bar{H}(Q) & \le\log\frac{\mu(S-B_{n})}{(\mu(B_{n}))^{2}}+4\\
 & \le\log\frac{\left(\frac{\sqrt{n}}{2}(\mu(B_{n}))^{1/n}+1\right)^{n}}{\mu(B_{n})}+4\\
 & \le n\log\left(\sqrt{\frac{\pi e}{2}}+1\right)-\log\mu(B_{n})+4,
\end{align*}
where the last inequality is because
\begin{align*}
\frac{\sqrt{n}}{2}(\mu(B_{n}))^{1/n} & =\frac{\sqrt{n}}{2}\left(\frac{\pi^{n/2}}{\Gamma(n/2+1)}\right)^{1/n}\\
 & \le\frac{\sqrt{n}}{2}\left(\frac{\pi^{n/2}}{\sqrt{\pi n}\left(\frac{n}{2e}\right)^{n/2}}\right)^{1/n}\\
 & =\frac{\sqrt{n}}{2}\sqrt{\frac{2\pi e}{n}}\left(\frac{1}{\sqrt{\pi n}}\right)^{1/n}\\
 & \le\sqrt{\frac{\pi e}{2}}.
\end{align*}

\section{Proof of Theorem \ref{thm:ub_on_ball}\label{subsec:thm_ball_pf}}

Let $\mathbf{G}$ be the generator matrix of $Q$, $T$ be a preimage basic cell of $Q$, 
and $\mathcal{Q} := \{\mathbf{q} = Q(\mathbf{x}):\, \mathbf{x} \in T\}$ be the image of $Q$ (which is finite or countable). 
For $\mathbf{x} \sim \mathrm{Unif}(T)$, we have
\begin{equation} \label{eq:converseEntropy}
H(Q(\mathbf{x})) = \mathbb{E}\left[- \log \frac{\mu(Q^{-1}(Q(\mathbf{x})))}{|\det \mathbf{G}|} \right] = -\sum_{\mathbf{q}\in \mathcal{Q}} \frac{\mu(Q^{-1}(\mathbf{q}))}{|\det \mathbf{G}|} \log \frac{\mu(Q^{-1}(\mathbf{q}))}{|\det \mathbf{G}|}.
\end{equation}

Since $\mathbf{x} - Q(\mathbf{x}) \in B_n$, we have $\mathbf{x} - \mathbf{q} \in B_n$ whenever $\mathbf{x} \in Q^{-1}(\mathbf{q})$, implying that the quantization cell $Q^{-1}(\mathbf{q}) \subseteq  B_n + \mathbf{q}$ is a subset of the unit ball centered at $\mathbf{q}$. This gives $\mu(Q^{-1}(\mathbf{q})) \le \mu(B_n)$. We are interested in quantization cells that occupy a large portion of the ball. Fix $0 < \gamma \leq 1$, and let 
\[
\tilde{\mathcal{Q}} := \big\{\mathbf{q}\in \mathcal{Q}:\, \mu(Q^{-1}(\mathbf{q})) \ge \gamma \mu(B_n)\big\}
\] 
be the set of reconstruction points such that the volumes of the quantization cells are at least $\gamma \mu(B_n)$ and let $k$ be the size of $\tilde{\mathcal{Q}}$, i.e., $k = |\tilde{\mathcal{Q}}|$.

We first prove the following claim on the intersection of two balls: for any $\mathbf{x}_1,\mathbf{x}_2 \in \mathbb{R}^n$ with $\Vert \mathbf{x}_1-\mathbf{x}_2 \Vert = 2d$, $0\le d \le 1$, we have
\[
\mu((B_n + \mathbf{x}_1) \cap (B_n + \mathbf{x}_2)) \ge \frac{2}{n}(1-d)(1-d^2)^{(n-1)/2} \mu(B_{n-1}).
\]
To prove this claim, we assume $\mathbf{x}_1=(-d,0,\ldots,0)$ and $\mathbf{x}_2=(d,0,\ldots,0)$ without loss of generality. We have
\begin{align*}
& \mu((B_n + \mathbf{x}_1) \cap (B_n + \mathbf{x}_2)) \\
& = \int_{-\infty}^{\infty} \mu\big((B_n + \mathbf{x}_1) \cap (B_n + \mathbf{x}_2) \cap \{x\} \times \mathbb{R}^{n-1}\big) \,\mathrm{d}x \\
& = \int_{-(1-d)}^{1-d} \mu(B_{n-1}) \big(1 - (|x|+d)^2\big)^{(n-1)/2} \,\mathrm{d}x \\
& \stackrel{(a)}{\ge} \int_{-(1-d)}^{1-d} \mu(B_{n-1}) \left(\frac{1 - d - |x|}{1-d} \sqrt{1-d^2}\right)^{n-1} \,\mathrm{d}x \\
& = 2\mu(B_{n-1}) \int_{0}^{1-d}  \left(\frac{1 - d - x}{1-d} \sqrt{1-d^2}\right)^{n-1} \,\mathrm{d}x \\ 
%by change of variable u = 1-d-x and u is dummy
& = 2\mu(B_{n-1}) \left(\frac{\sqrt{1-d^2}}{1-d} \right)^{n-1} \int_{0}^{1-d}  x^{n-1} \,\mathrm{d}x \\
& = 2\mu(B_{n-1}) \left(1-d^2 \right)^{(n-1)/2} \frac{1-d}{n},
\end{align*}
where (a) is because $\sqrt{1-y^2} \ge \sqrt{1-d^2} (1-y)/(1-d)$ for $d \le y \le 1$ due to the concavity of $\sqrt{1-y^2}$.

Let 
\[
D = \min_{\mathbf{q}_1,\mathbf{q}_2 \in \tilde{\mathcal{Q}} + \mathbf{G}\mathbb{Z}^n,\, \mathbf{q}_1 \neq \mathbf{q}_2} \frac{\Vert\mathbf{q}_1 - \mathbf{q}_2 \Vert}{2}
\]
be half the minimum distance between two points in $\tilde{\mathcal{Q}}$ and its lattice-translated copies. Since 
\begin{align*}
    &\mu((B_n + \mathbf{q}_1) \cup (B_n + \mathbf{q}_2)) \\
    &\ge \mu(Q^{-1}(\mathbf{q}_1)) + \mu(Q^{-1}(\mathbf{q}_2)) \\
    &\ge 2\gamma \mu(B_n)
\end{align*} 
for distinct $\mathbf{q}_1,\mathbf{q}_2 \in \tilde{\mathcal{Q}} + \mathbf{G}\mathbb{Z}^n$, by the claim, we have
%\[2\mu(B_n) = \mu((B_n + \mathbf{q}_1) \cup (B_n + \mathbf{q}_2)) + \mu((B_n + \mathbf{q}_1) \cap (B_n + \mathbf{q}_2)) \ge 2\gamma \mu(B_n) + \mu((B_n + \mathbf{q}_1) \cap (B_n + \mathbf{q}_2))\]
% Let $2d$ be the distance between the centers of two unit balls where $0 < d \leq 1$. 
% \textcolor{red}{Then, it is easy to see that \textcolor{blue}{(maybe draw a figure)} the difference between the volume of two unit balls and the volume of two quantization cells with the smallest volume $\gamma \mu(B_{n})$ is at least as large as the volume of two symmetric $n$-dimensional cones inscribed inside the lens in the overlapping region (if the overlapping region is nonempty and has nonzero volume) of the unit balls}, and thus,
\begin{align*}
2 (1-\gamma) \mu(B_n) & \ge \mu((B_n + \mathbf{q}_1) \cap (B_n + \mathbf{q}_2))\\
&\ge \frac{2}{n}(1-D)(1-D^2)^{(n-1)/2} \mu(B_{n-1}) \\
& \ge \frac{2}{n}(1-D)^{(n+1)/2} \mu(B_{n-1}),
\end{align*}
which implies that
\begin{align}
D  &\ge 1- \left(\frac{n (1-\gamma) \mu(B_n)}{\mu(B_{n-1})}\right)^{2/(n+1)} \nonumber\\
& = 1- \left(n (1-\gamma)\sqrt{\pi}\frac{\Gamma(n/2+1/2)}{\Gamma(n/2+1)}\right)^{2/(n+1)} \nonumber\\
& \stackrel{(b)}{\ge}  1- \left(\frac{n \sqrt{2\pi}(1-\gamma)}{\sqrt{n}}\right)^{2/(n+1)}  \nonumber\\
&= 1- \left( \sqrt{2\pi n}(1-\gamma)\right)^{2/(n+1)},
\label{eq:distance}
\end{align}
where (b) is by Wendel's inequality~\cite{Wendel1948Note,qi2009bounds}.
% \[\frac{\Gamma(n/2+1/2)}{\Gamma(n/2+1)} = \frac{1}{(\Gamma(n/2+1)/\Gamma(n/2+1/2)} \leq (n/2)^{1/2-1} = (n/2)^{-1/2}\]
%  because
% \[\frac{\Gamma(n/2+1)}{\Gamma(n/2+1/2)} \geq \frac{1}{x^{s-1}}\] 
%  with $x = n/2$
% \[
% \frac{n/2+1/2}{\sqrt{n/2+1}}
% \]

Note that the balls $DB_n + \mathbf{q}$ for $\mathbf{q} \in \tilde{\mathcal{Q}} + \mathbf{G}\mathbb{Z}^n$ form a sphere packing in $\mathbb{R}^n$.
Let $\eta_n$ be the optimal density of sphere packing over $\mathbb{R}^n$, 
 and let $\ell = |\det \mathbf{G}|/\mu(B_n)$.
Then we have
\begin{align*}
\eta_n &\ge \frac{k \mu(DB_n)}{|\det \mathbf{G}|} \\
&=\frac{D^n k \mu(B_n)}{|\det \mathbf{G}|} \\
& =  \frac{D^n k}{\ell},
\end{align*}
which implies that
\begin{align} \label{eq:density}
\frac{k}{\ell} \le \frac{\eta_n}{D^n }.
\end{align}
Hence, from \eqref{eq:converseEntropy}, we have
\begin{align}
H(Q(\mathbf{x})) &= \mathbb{E}\left[- \log \frac{\mu(Q^{-1}(Q(\mathbf{x})))}{|\det \mathbf{G}|} \right] \nonumber\\
&= -\sum_{\mathbf{q}\in \mathcal{Q}} \frac{\mu(Q^{-1}(\mathbf{q}))}{|\det \mathbf{G}|} \log \frac{\mu(Q^{-1}(\mathbf{q}))}{|\det \mathbf{G}|} \nonumber\\
&= \sum_{\mathbf{q}:\mu(Q^{-1}(\mathbf{q})) \geq \gamma \mu(B_{n})} \frac{\mu(Q^{-1}(\mathbf{q}))}{|\det \mathbf{G}|} \log \frac{|\det \mathbf{G}|}{\mu(Q^{-1}(\mathbf{q}))} \nonumber\\ &\quad+\sum_{\mathbf{q}:\mu(Q^{-1}(\mathbf{q})) < \gamma \mu(B_{n})} \frac{\mu(Q^{-1}(\mathbf{q}))}{|\det \mathbf{G}|} \log \frac{|\det \mathbf{G}|}{\mu(Q^{-1}(\mathbf{q}))} \nonumber\\
&\ge\log \frac{|\det \mathbf{G}|}{\mu(B_n)} \sum_{\mathbf{q}:\mu(Q^{-1}(\mathbf{q})) \geq \gamma \mu(B_{n})} \frac{\mu(Q^{-1}(\mathbf{q}))}{|\det \mathbf{G}|}  \nonumber\\ &\quad+\log \frac{|\det \mathbf{G}|}{\gamma\mu(B_n)}\sum_{\mathbf{q}:\mu(Q^{-1}(\mathbf{q})) < \gamma \mu(B_{n})} \frac{\mu(Q^{-1}(\mathbf{q}))}{|\det \mathbf{G}|}  \nonumber\\
& = \log \frac{\ell}{\gamma} - \bigg(\log \frac{1}{\gamma}\bigg)\bigg(\sum_{\mathbf{q}:\mu(Q^{-1}(\mathbf{q})) \geq \gamma \mu(B_{n})} \frac{\mu(Q^{-1}(\mathbf{q}))}{|\det \mathbf{G}|}\bigg) \nonumber\\
& \ge \log \frac{\ell}{\gamma} - \bigg(\log \frac{1}{\gamma}\bigg)\bigg(\sum_{\mathbf{q}:\mu(Q^{-1}(\mathbf{q})) \geq \gamma \mu(B_{n})} \frac{\mu(B_n)}{|\det \mathbf{G}|}\bigg) \nonumber\\
& = \log \frac{\ell}{\gamma} - \bigg(\log \frac{1}{\gamma}\bigg)\frac{k}{\ell} \nonumber\\
%& \stackrel{(b)}{\ge} \frac{k}{\ell} \log \ell + \left(1- \frac{k}{\ell} \right) \log \frac{\ell}{\gamma} \nonumber\\
& = \log \ell + \left(1- \frac{k}{\ell} \right) \log \frac{1}{\gamma} \nonumber\\
& \stackrel{(c)}{\ge} \log \ell + \left(1- \frac{\eta_n}{D^n} \right) \log \frac{1}{\gamma} \nonumber\\
& \stackrel{(d)}{\ge} \log \ell + \left(1- \frac{\eta_n}{D^n} \right) (1- \gamma) \log e \nonumber\\
& \stackrel{(e)}{\ge} \log \ell + \left(1- \frac{\eta_n}{\left(1- \left(\sqrt{2\pi n}(1-\gamma)\right)^{2/(n+1)}\right)^n} \right) (1- \gamma) \log e \label{eq:HQ_lowerbound}
\end{align}
where (c) is by~\eqref{eq:density}; (d) is because of the inequality $\ln{a} \le a - 1 $ for any $a > 0$ and a change of base; 
% 1 - a \leq \ln(1/gamma) = log(1/gamma)/log(e)
and (e) is by~\eqref{eq:distance}.

Now, let $\bar{\gamma} = 1 - \gamma$ and consider the real-valued function $f$ defined by
\begin{equation}
f(\bar{\gamma}) := \left(1- \frac{\eta_n }{\left(1- \left(\sqrt{2\pi n}\bar{\gamma} \right)^{2/(n+1)}\right)^n} \right). 
\end{equation}
Notice that maximizing the lower bound~\eqref{eq:HQ_lowerbound} on $H(Q(\mathbf{x}))$ over $0<\gamma \le 1$ is equivalent to maximizing the function $\bar{\gamma} f(\bar{\gamma})$ over $0 \leq \bar{\gamma} < 1$. 
It can be checked that $f(0) = 1-\eta_{n}$ (i.e., when $\gamma = 1$) and $f(\bar{\gamma})$ is monotonically decreasing over the interval $[0,1)$.
Moreover, when $f(\bar{\gamma}_{0})=0$, we have 
\[
\left(\sqrt{2\pi n}\bar{\gamma}_{0}\right)^{2/(n+1)}= 1-\eta_{n}^{1/n},
\]
which implies that
\[\bar{\gamma}_{0}=\frac{(1-\eta_{n}^{1/n})^{(n+1)/2}}{\sqrt{2\pi n}}.\]
Therefore, the maximum of the function $\bar{\gamma} f(\bar{\gamma})$ occurs in the interval $[0,\bar{\gamma}_{0}]$. Although the bound is the tightest when we consider the $\bar{\gamma}$ attaining the maximum, for the sake of simplicity, we simply take $\bar{\gamma}_{*} = \bar{\gamma}_{0}/2$, which gives
\begin{align}
H(Q(\mathbf{x})) &\ge \log \ell + \left(1- \frac{\eta_n}{\left(1- \left(\frac{(1-\eta_{n}^{1/n})^{(n+1)/2}}{2}\right)^{2/(n+1)}\right)^n} \right) \Bigg(\frac{(1-\eta_{n}^{1/n})^{(n+1)/2}}{2\sqrt{2\pi n}}\Bigg) \log e \nonumber\\ 
&= \log \ell + \left(1- \frac{\eta_n}{\left(1- \frac{(1-\eta_{n}^{1/n})}{4^{1/(n+1)}}\right)^n} \right) \Bigg(\frac{(1-\eta_{n}^{1/n})^{(n+1)/2}}{2\sqrt{2\pi n}}\Bigg) \log e ,\\ \nonumber
&= \log \ell + \left(1- \frac{4^{n/(n+1)}\eta_n}{\left(4^{1/(n+1)}- (1-\eta_{n}^{1/n})\right)^n} \right) \Bigg(\frac{(1-\eta_{n}^{1/n})^{(n+1)/2}}{2\sqrt{2\pi n}}\Bigg) \log e ,\\ \nonumber
\end{align}
and 
\begin{equation}
    \bar{H}(Q(\mathbf{x})) \geq \left(1- \frac{4^{n/(n+1)}\eta_n}{\left(4^{1/(n+1)}- (1-\eta_{n}^{1/n})\right)^n} \right) \Bigg(\frac{(1-\eta_{n}^{1/n})^{(n+1)/2}}{2\sqrt{2\pi n}}\Bigg) \log e - \log \mu(B_{n}). 
\end{equation}

\section{Calculations in Example \ref{exa:disk}\label{subsec:disk_pf}}

We now study a shift-periodic quantizer over $\mathbb{R}^2$ with an error distribution that is uniform over the unit disk. Consider the unit disk $B_2 \subseteq \mathbb{R}^2$ centered at the origin, and a basic cell $S$ given by the fundamental Voronoi region (which is a regular hexagon) induced by the hexagonal lattice $\Lambda_{1}$ generated by the basis vectors $\{\mathbf{g}_{1} = \alpha (0,2)^{T}, \mathbf{g}_{2} = \alpha(\sqrt{3},1)^{T}\}$, where $\alpha = \sqrt{\frac{\pi}{2\sqrt{3}}}$ is a scaling factor which makes $\mu(S) = |\det(\mathbf{G}_{1})| = 2 \sqrt{3} \alpha^2 = \pi$ equals the area of $B_2$.
We consider the hexagonal lattice instead of a square lattice since a hexagon is closer to a disk, and hence the $\mu(S\backslash A)$ term in Theorem~\ref{thm:quantization} will be smaller.
% Note that the disk $B_2$ has area $\pi$.
% Since the volume of any basic cell is invariant to the lattice partition and is given by the lattice determinant, i.e., the absolute value of the determinant of the generator matrix $\mathbf{G}_{1}=(\mathbf{g}_{1} \; \mathbf{g}_{2})$ formed by the basis vectors, we have 
% \begin{align*}
% \mu(S) = |\det(\mathbf{G}_{1})| = 2 \sqrt{3}.
% \end{align*}
% A generator matrix $\mathbf{G}_{1}'$ for the hexagonal lattice $\Lambda_{1}'$ that induces the Voronoi region $S'$ of area $\pi$ is given as follows:
% \begin{align*}
%     \mathbf{G}_{1}' = \alpha \cdot \mathbf{G}_{1} = 
% \alpha \cdot
% \begin{pmatrix}
% 0 & \sqrt{3} \\
% 2 & 1
% \end{pmatrix},
% \end{align*}
% where $\alpha = \sqrt{\frac{\pi}{2\sqrt{3}}}$.
It follows that the apothem $a$ of $S$ (or equivalently the length of the face-determining points of $\Lambda_1$) and the circumradius $R$ of $S$ are $a=\sqrt{\frac{\pi}{2\sqrt{3}}}$ and $R=\frac{a}{\cos \frac{\pi}{6}} = \sqrt{\frac{2\pi}{3 \sqrt{3}}}$, respectively.
By a simple geometric argument, we can derive that the area of $S \backslash B_2$ is given by 
\begin{align*}
    \mu(S\backslash B_2) = 6\left(\frac{\theta}{2}-\frac{\sin \theta}{2}\right) = 3\left(\theta - \sin \theta \right),
\end{align*}
where $\theta = 2 \arccos\left(a\right) = 2 \arccos \left(\sqrt{\frac{\pi}{2\sqrt{3}}}\right)$.
Therefore, the normalized entropy of the shift-periodic quantizer $Q$ with $S=S$ and $A=B_2$ constructed as in Theorem~\ref{thm:quantization} (second part) is upper-bounded by 
\begin{align*}
& H_{b}\left(\frac{\mu(S\backslash B_2)}{\mu(B_2)}\right) \\
&\;\; +\frac{\mu(S\backslash B_2)}{\mu(B_2)}\left(\!\log\frac{\mu((S\backslash B_2)-(B_2\backslash S))}{\mu(S\backslash B_2)}+4\!\right)-\log\mu(B_2) \\
&\le H_{b}\left(\frac{\mu(S\backslash B_2)}{\mu(B_2)}\right)+\frac{\mu(S\backslash B_2)}{\mu(B_2)}\left(\log\frac{\mu((1+R)B_2)}{\mu(S\backslash B_2)}+4\right)\\
&\;\;\;-\log\mu(B_2) \\ 
&\le -1.01666 \;\mathrm{bits}. % the exact value is -1.0166631245419808, rounding up to -1.01666
\end{align*}
% where $D'$ is a disk centered at origin with radius $1+R$ and contains the region $(S\backslash B_2)-(B_2\backslash S)$.

\section{Calculations in Example \ref{exa:ball}\label{subsec:ball_pf}}

We then study a shift-periodic quantizer over $\mathbb{R}^3$ with an error distribution that is uniform over the unit ball.
Consider the unit ball $B_3$ centered at the origin, and a basic cell $S_{1}$ given by the fundamental Voronoi region (which is a rhombic dodecahedron) induced by the face-centered cubic lattice $\Lambda_{1}$ generated by the basis vectors $\{\mathbf{g}_{1} = \beta(-1,-1,0)^{T}, \mathbf{g}_{2} = \beta(1,-1,0)^{T}, \mathbf{g}_{3} = \beta(0,1,-1)^{T}\}$, where $\beta = \sqrt[3]{\frac{2\pi}{3}}$ is a scaling factor which makes $\mu(S_{1}) = |\det(\mathbf{G}_{1})| = 2 \beta^{3} = \frac{4\pi}{3}$ equals the volume of $B_3$.
The 8 vertices of $S_{1}$ where 3 faces meet at the obtuse angle are $\beta(\pm \frac{1}{2}, \pm \frac{1}{2}, \pm \frac{1}{2})$ while the 6 vertices of $S_{1}$ where 4 faces meet at the acute angle are $\beta(\pm 1,0,0)$, $\beta(0,\pm 1,0)$, and $\beta( 0,0,\pm 1)$, resulting in a total of 14 vertices.
By using numerical method\footnote{For ease of computation, we approximate a ball $B_3$ by a polytope $P$.
% by the intersection of finitely many half-spaces (i.e, $H$-representation), say $H_{1}$. 
% Let $H_{2}$ be the $H$-representation of a parallelohedron $S$. 
Let $S$ be a polytope.
We numerically approximate the volume of the intersection $B_3 \cap S$ as the volume of $P \cap S$. Then, $\mu(B_3\backslash S)=\mu(B_3)-\mu(B_3 \cap S)\approx \mu(B_3)-\mu(P \cap S)$.}, we can compute that the volume of $B_3 \backslash S_{1}$ (or equivalently $S_{1} \backslash B_3$) is upper-bounded by 0.33153.
% Note: We approximate the region of the ball by inner polygon, so we are underestimate the volume of intersection. Thus, we overestimate the volume of leftover.
Therefore, the normalized entropy of the shift-periodic quantizer $Q$ with $S=S_{1}$ and $A=B_3$ constructed as in Theorem~\ref{thm:quantization} (second part) is upper-bounded by 
\begin{align*}
& H_{b}\left(\frac{\mu(S_{1}\backslash B_3)}{\mu(B_3)}\right) \\
&\;+\frac{\mu(S_{1}\backslash B_3)}{\mu(B_3)}\left(\!\log\frac{\mu((S_{1}\backslash B_3)-(B_3\backslash S_{1}))}{\mu(S_{1}\backslash B_3)}\!+4\!\right)\!-\log\mu(B_3) \\
&\le H_{b}\left(\frac{\mu(S_{1}\backslash B_3)}{\mu(B_3)}\right)+\frac{\mu(S_{1}\backslash B_3)}{\mu(B_3)}\left(\log\frac{\mu((1+\beta)B_3)}{\mu(S_{1}\backslash B_3)}+4\right) \\
&\;\;\;-\log\mu(B_3) \\ 
&\le -0.77892 \;\mathrm{bits}. % the calculated value is -0.7789232226667471
\end{align*}
% where $B'$ is a ball centered at origin with radius $1+\beta$ and contains the region $(S_{1}\backslash B)-(B\backslash S_{1})$.

For another example in the 3-dimensional Euclidean space, consider again the ball $B_3$ with unit radius centered at the origin and the fundamental Voronoi region $S_{2}$ (which is a truncated octahedron) induced by the body-centered cubic lattice
% \footnote{A body-centered cubic lattice is dual to a face-centered cubic lattice.}
$\Lambda_{3}$ generated by the basis vectors $\{\mathbf{g}_{1} = \gamma(2, 0, 0)^{T}, \mathbf{g}_{2} = \gamma(0, 2, 0)^{T}, \mathbf{g}_{3} = \gamma(1, 1, 1)^{T}\}$, where $\gamma = \sqrt[3]{\frac{\pi}{3}}$ is a scaling factor which makes $\mu(S_{2}) = |\det(\mathbf{G}_{2})| = 4\gamma^{3} = \frac{4\pi}{3}$ equals the volume of $B_3$.
All the permutations of $\gamma(0, \pm \frac{1}{2}, \pm 1)$ are the vertices of $S_{2}$, resulting in a total of 24 vertices.
Similarly, by using numerical method, we can compute that the volume of $B_3 \backslash S_{2}$ (or equivalently $S_{2} \backslash B_3$) is upper-bounded by 0.35063.
Therefore, the normalized entropy of the shift-periodic quantizer $Q$ with $S=S_{2}$ and $A=B_3$ constructed as in Theorem~\ref{thm:quantization} (second part) is upper-bounded by 
\begin{align*}
& H_{b}\left(\frac{\mu(S_{2}\backslash B_3)}{\mu(B_3)}\right)\\
&\; +\frac{\mu(S_{2}\backslash B_3)}{\mu(B_3)}\left(\!\log\frac{\mu((S_{2}\backslash B_3)-(B_3\backslash S_{2}))}{\mu(S_{2}\backslash B_3)}+4\!\right)\!-\log\mu(B_3) \\
&\le H_{b}\left(\frac{\mu(S_{2}\backslash B_3)}{\mu(B_3)}\right) \\
&\;\;\;+\frac{\mu(S_{2}\backslash B_3)}{\mu(B_3)}\left(\log\frac{\mu((1+\frac{\sqrt{5}}{2}\gamma)B_3)}{\mu(S_{2}\backslash B_3)}+4\right)-\log\mu(B_3) \\ 
&\le -0.74221 \;\mathrm{bits}. % the calculated value is -0.7422100677101326
\end{align*}

\fi

\end{document}